\documentclass[11pt,letterpaper,usenames,dvipsnames]{article}
\usepackage{jheppub}

\allowdisplaybreaks[2]  
\usepackage{pifont}
\usepackage{bm} 
\usepackage{dsfont} 
\usepackage{ytableau}   
\usepackage{float}
\usepackage{hhline}
\usepackage{mathtools}

\usepackage{verbatim}
\usepackage{tikz}
\usetikzlibrary{arrows,snakes,shapes.arrows,decorations.markings}
     \tikzset{>=triangle 90}
     \tikzstyle{gr}=[draw,circle,green!50!black,fill=green!50!black,scale=.6]
     \tikzstyle{Bl}=[draw,circle,blue,scale=.7]
     \tikzstyle{R}=[draw,circle,fill=red,scale=.7]
     \tikzstyle{bl}=[draw,circle,fill=black,scale=.2]
     \tikzstyle{bbc}=[draw,circle,fill=black,scale=.75]
     \tikzstyle{bbcs}=[draw,circle,fill=black,scale=.5]
     \tikzstyle{rc}=[circle,fill=red,scale=.6]
     \tikzstyle{wc}=[draw,circle,scale=.75]
\usepackage{array} 
\usepackage{multirow} 
\usepackage{colortbl} 
\usepackage{stfloats}  
\usepackage{cellspace}
\usepackage{xcolor}   
\def\blue#1{{\color{blue}{#1}}}
\def\green#1{{\color{black!25!green}{#1}}}

\def\red#1{{\color{red}{#1}}}
\def\grey#1{{\color{black!40}{#1}}}

\def\ccy{\cellcolor{yellow!70}}

\def\rcy{\rowcolor{black!25!yellow!10}}
\def\rco{\rowcolor{black!25!orange!15}}

\def\csc{\scriptstyle}
\def\yd{\ydiagram}

\usepackage{url} 

\newcommand{\beq}{\begin{equation}}
\newcommand{\eeq}{\end{equation}}
\newcommand{\xdasharrow}[2][->]{
\tikz[baseline=-\the\dimexpr\fontdimen22\textfont2\relax]{
\node[anchor=south,font=\scriptsize, inner ysep=1.5pt,outer xsep=2.2pt](x){#2};
\draw[shorten <=3.4pt,shorten >=3.4pt,dashed,#1](x.south west)--(x.south east);
}
}

\def\bar{\overline}
\def\til{\widetilde}
\def\hat{\widehat}

\def\^{\wedge}

\def\Aut{\mathop{\rm Aut}}
\def\Inn{\mathop{\rm Inn}}
\def\Out{\mathop{\rm Out}}

\def\SU{\mathop{\rm su}}

\def\SO{\mathop{\rm so}}
\def\SL{\mathop{\rm SL}}
 
\def\Sp{\mathop{\rm sp}}

\def\C{\mathbb{C}} 
\def\H{\mathbb{H}}

\def\R{\mathbb{R}} 
\def\Z{\mathbb{Z}} 
\def\ff{{\mathfrak f}}

\def\cB{{\mathcal B}}
\def\cBh{{\hat\cB}}

\def\cE{{\mathcal E}}

\def\cN{{\mathcal N}}
\def\cO{{\mathcal O}}

\def\a{{\alpha}}

\def\G{{\Gamma}}
\def\d{{\delta}}

\def\D{{\Delta}}
\def\e{{\epsilon}}

\def\th{{\theta}}

\def\l{{\lambda}}
\def\L{{\Lambda}}
\def\m{{\mu}}

\def\n{{\nu}}

\def\r{{\rho}}
\def\s{{\sigma}}

\def\t{{\tau}}

\def\f{{\phi}}
\def\vf{{\varphi}}

\def\ssqt{{\!\scriptscriptstyle\sqrt2}}
\def\Chi{\raisebox{\depth}{$\csc\chi$}}
\def\cda{{\quad\grey\bf\downarrow}}
\def\cvl{{\quad\hspace{1.5pt}\blue\bf\shortmid}}

\title{4d $\cN$=2 theories with disconnected gauge groups}
\author{Philip C. Argyres,}
\author{and Mario Martone}
\affiliation{University of Cincinnati,
Physics Department, PO Box 210011, Cincinnati OH 45221}
\emailAdd{philip.argyres@gmail.com}
\emailAdd{martonmo@ucmail.uc.edu}

\abstract{
In this paper we present a beautifully consistent web of evidence for the existence of interacting 4d rank-1 $\cN=2$ SCFTs obtained from gauging discrete subgroups of global symmetries of other existing 4d rank-1 $\cN=2$ SCFTs.  The global symmetries that can be gauged involve a non-trivial combination of discrete subgroups of the $U(1)_R$, low-energy EM duality group $SL(2,\Z)$, and the outer automorphism group of the flavor symmetry algebra, Out($F$).

The theories that we construct are remarkable in many ways: (i) two of them have exceptional $F_4$ and $G_2$ flavor groups; (ii) they substantially complete the picture of the landscape of rank-1 $\cN=2$ SCFTs as they realize all but one of the remaining consistent rank-1 Seiberg-Witten geometries that we previously constructed but were not associated to known SCFTs; and (iii) some of them have enlarged $\cN=3$ SUSY, and have not been previously constructed.  They are also examples of SCFTs which violate the Shapere-Tachikawa relation between the conformal central charges and the scaling dimension of the Coulomb branch vev.  We propose a modification of the formulas computing these central charges from the topologically twisted Coulomb branch partition function which correctly compute them for discretely gauged theories.
}
 
\begin{document}
\maketitle

\section{Introduction}

In this paper we investigate all possible discrete symmetries which can be gauged in 4d rank-1 SCFTs while preserving $\cN=2$ supersymmetry.  The idea of gauging a discrete symmetry was first introduced a long time ago in the context of field theories on a lattice \cite{Wegner:1984qt} and then later extended to the continuum case \cite{Krauss:1988zc, Banks:1989ag}.  
Gauging of discrete symmetries in the context of 4d superconformal field theories (SCFTs) was recently discussed in \cite{gr1512} and \cite{Aharony:2016kai}, whose ideas have strongly influenced this paper.


We will show, on the one hand, that intricate consistency conditions need to be satisfied for the existence of a discretely gauged version of a rank-1 $\cN=2$ SCFT, and, on the other hand, that these conditions have a rich set of solutions, enabling us to construct many new theories.  Some of these theories have exceptional flavor groups --- in particular $F_4$ and $G_2$ --- or extended $\cN=3$ supersymmetry.  Our results are summarized in table \ref{tab1}.

\begin{table}[ht]
\centering\footnotesize
$\def\arraystretch{0.6}
\begin{array}{llllllllll|cccc}
\multicolumn{10}{c|}{\text{\bf Discrete gauge group action on the Coulomb Branch:}} &
\multicolumn{3}{l}{\text{\bf \ CFT data:}}
\\[1mm]
\multicolumn{1}{c}{\text{\bf parent}}
& \multicolumn{1}{c}{\bf\Z_2} 
&\multicolumn{1}{c}{\green{\bf\til \Z_2}} 
&\multicolumn{1}{c}{\bf\Z_3} 
& \multicolumn{1}{c}{\green{\bf\til \Z_3}} 
& \multicolumn{1}{c}{\bf\Z_4} 
& \multicolumn{1}{c}{\green{\bf\til \Z_4}} 
& \multicolumn{1}{c}{\bf\Z_5} 
& \multicolumn{1}{c}{\bf\Z_6} 
& \multicolumn{1}{c|}{\green{\bf\til \Z_6}} 
&\csc\boldsymbol{k_F}\ &\csc\boldsymbol{24 a} 
&\csc\boldsymbol{12 c} &\csc\boldsymbol h\\[1.5mm]
\hline
&&&&&&&&&\\[-1.5mm]
\csc [II^*,E_8] &&&&&&&&&&
12& 95 & 62 & 0\\
\cda &&&&&&&&&\\
\csc [III^*,E_7] &&&&&&&&&& 
8&59 & 38 & 0\\
\cda &&&&&&&&&\\
\csc [IV^*,E_6] & \csc [II^*,F_4] &&&&&&&&&
6& 41 & 26 & 0\\
\cda & \cda&&&&&&&&\\
\rco \csc [I_0^*,D_4\Chi_0] & \csc [III^*,B_3] & & \csc [II^*,G_2]& &&&&&&
4&23 & 14 & 0\\
\rco \cda &\cda&&\cvl&&&&&&&&&&\\
\csc [IV,A_2\Chi_{\frac12}] & \csc [IV^*, A_2] &&\cvl && \csc [II^*,B_1] &&&&&
3&14 & 8 & 0\\
\cda &\cvl&&\cda&&\cvl&&&&&&\\
\csc [III,A_1\Chi_{\frac23}] &\cvl&& \csc [III^*,A_1] 
&&\cvl&&&&&
\frac83&11 & 6 & 0\\
\cda &\cvl&&\cvl&&\cvl&&&&&&\\
\csc [II,\Chi_{\frac45}] &\cvl&&\cvl&&\cvl&& 
\csc \!\!\!\!\red{[II^*,\varnothing]}\!\!\!\! \ \ &&&
-& \frac{43}{5} & \frac{22}{5} & 0\\
\cda &\cda&&\cda&&\cda&&&&&&\\
\rcy \csc [I_0,\varnothing] & \csc [I_0^*,\varnothing] &&
\csc [IV^*_1,\varnothing] && \csc [III^*,\varnothing] 
&&& \csc [II^*,\varnothing] &&
-& 5 & 2 & 0\\[1.5mm]
\hline
&&&&&&&&&\\[-1.5mm]
\csc [II^*,C_5] &&&&&&&&&&
7& 82 & 49 & 5\\
\cda &&&&&&&&&&&\\
\csc [III^*,C_3C_1] &&&&&&&&&&
\csc (5,8)& 50 & 29 & 3\\
\cda &&&&&&&&&&&\\
\csc [IV^*,C_2U_1] & \csc [II^*,C_2] &&&&&&&&&
\csc (4,?)& 34 & 19 & 2\\
\cda &\cda&&&&&&&&&&\\
\rco \csc \blue{[I_0^*,C_1\Chi_0]} & \csc [III^*,C_1] & \csc\green{[III^*,U_1{\!\rtimes}\Z_2]} &&&&&&&&
3& 18 & 9 & 1\\
\rco\cda &\cda&\cda&&&&&&&&&&&\\
\rcy \csc [I_4,U_1] & \csc [I_2^*,\varnothing] & \csc [I_2^*,\varnothing]&&&&&&&& 
1 & 6 & 3 & 0\\[1.5mm]
\hline
&&&&&&&&&\\[-1.5mm]
\csc [II^*,A_3{\!\rtimes}\Z_2] &&&&&&&&&&
14& 75 & 42 & 4\\
\cda &&&&&&&&&&&\\
\csc [III^*,A_1U_1{\!\rtimes}\Z_2] \!\!\!\!\!\!\!\!\!\!\!
&&&&&&&&&&
\csc (10,?)\!\!\!\!\!& 45 & 24 & 2\\
\cda &&&&&&&&&&&\\
\csc \green{[IV^*,U_1]} & \csc \red{[II^*,\varnothing]} &&&&&&&&&
5 & 30 & 15 & 1\\
\cda &&&&&&&&&&&\\
\rcy \csc [I_1^*,\varnothing] &&&&&&&&&&
-& 17 & 8 & 0\\[1.5mm]
\hline
&&&&&&&&&\\[-1.5mm]
\csc [II^*,A_2{\!\rtimes}\Z_2]\! &&&&&&&&&&
14& 71 & 38 & 3\\
\cda &&&&&&&&&&&\\
\csc \green{[III^*,U_1{\!\rtimes}\Z_2]}\!\!\! &&&&&&&&&&
7 &42 & 21 & 1\\
\cda &&&&&&&&&&&\\
\csc [IV^*_1,\varnothing] &&&&&&&&&&
- &\frac{55}{2} & \frac{25}{2} & 0\\[1.5mm]
\hline
\rco&&&&&&&&&&&&&\\[-1.5mm]
\rco \csc \blue{[I_0^*,C_1\Chi_0]} & \csc [III^*,C_1]& \csc \green{[III^*,U_1{\!\rtimes}\Z_2]}\ 
& \csc [II^*,C_1]& 
\csc \green{[II^*, U_1{\!\rtimes}\Z_2]}\!\!\! &&&&&&
3& 18 & 9 & 1\\
\rco\cda &\cda&\cda&\cvl&\cvl&&&&&&&&&\\
\rcy \csc [I_2,U_1] & \csc [I_1^*,\varnothing]& \csc [I_1^*,\varnothing] &\cvl  &\cvl &&&&&&
1&6 & 3 & 0\\
\rcy\cda &&&\cda&\cda&&&&&&&&&\\
\rcy \csc [I_0,\varnothing] &&& \csc [IV^*_\ssqt,\varnothing]& \csc [IV^*_\ssqt,\varnothing] \ \ &&&&&&
-& 5 & 2 & 0\\[1mm]
\hline
\rcy&&&&&&&&&&&\\[-2mm]
\rcy \csc \blue{[I_0,C_1\Chi_0]}&
\csc [I_0^*,\Chi_0]{\times}\H &
\csc \blue{[I_0^*,C_1\Chi_0]} &
\csc [IV^*_1,\varnothing]{\times}\H & 
\csc \green{[IV^*,U_1]} &
\csc [III^*,\varnothing]{\times}\H & 
\csc \green{[III^*,U_1{\!\rtimes}\Z_2]}\!\!\! & & 
\csc [II^*,\varnothing]{\times}\H &
\csc \green{[II^*,U_1{\!\rtimes}\Z_2]}\! & 1 & 6 & 3 & 1 \\
\rcy\cda&\cda&\cda&\cda&\cda&\cda&\cda&&\cda&\cda&&&&\\
\rcy \csc [I_0,\varnothing]& 
\csc [I_0^*,\varnothing] & \csc [I_0^*,\varnothing] &
\csc [IV^*_1,\varnothing] & \csc [IV^*_1,\varnothing] &
\csc [III^*,\varnothing] & \csc [III^*,\varnothing] &&
\csc [II^*,\varnothing] & \csc [II^*,\varnothing]& 
-&5 & 2 & 0\\
\end{array}$
\caption{Rank-1 $\cN{=}2$ SCFTs.  The notation is explained in the text; black entries have $\cN{=}2$ supersymmetry, green $\cN{=}3$, blue $\cN{=}4$, and the two red entries are somewhat more speculative --- i.e., there is little evidence from self-consistency checks for their existence.  
The vertical arrows denote some characteristic $\cN{=}2$ RG flows.  Darkly-shaded rows are lagrangian CFTs and lightly-shaded rows are IR-free or free theories.  The second-to-last row is a free $\cN{=}4$ vector multiplet and its discretely gauged versions.  The last four columns record the flavor ($k_F$) and conformal ($a$, $c$) central charges, and the quaternionic dimension ($h$) of the enhanced Coulomb branch fiber common to the theories in each row.  $\cN{=}4$ parent theories admit additional $\cN{=}3$-preserving discrete gaugings shown in the $\green{\til{\Z}_k}$ columns. \label{tab1}}
\end{table}

Gauging a discrete symmetry does not introduce any extra interactions.  Rather it simply acts as a superselection rule on the operator spectrum of the theory projecting out all operators which are not invariant under the gauged discrete symmetry.  
This means that gauging a discrete symmetry does not change the local dynamics of a theory, though it does change the spectrum of local and non-local operators.  For simplicity consider an operator $\cO$ which is odd under a $\Z_2$ symmetry:
$\cO(x)\xrightarrow{\Z_2}-\cO(x)$.
If this $\Z_2$ is gauged, the operator $\cO(x)$ is not a gauge-invariant local operator and so the state it creates from the vacuum, $|\cO(x)\rangle$, is projected out of the Hilbert space.  But $\cO(x)$ is not removed from the theory in the following sense.  Since a product of two $\cO$ operators is even under the $\Z_2$, $|\cO(x)\cO(y)\rangle$ will be part of the spectrum.  We can prepare a state arbitrarily close to $|\cO(x)\rangle$ by taking $y$ distant and space-like separated from $x$, thus leaving the local dynamics unchanged.  We will see in section \ref{sec:HB} how this is reflected in the structure of the Higgs branch of $\cN=2$ SCFT moduli spaces.
Also, because the local dynamics is unchanged, gauging a discrete symmetry does not modify the value of the conformal ($a$, $c$) and flavor ($k_F$) central charges from their values in the SCFT where the discrete symmetry is not gauged.

The discrete gauging operation turns out to organize the classification of 4d rank-1 $\cN=2$ SCFTs \cite{paper1, paper2, allm1602, Argyres:2016xmc} in a striking way.  That classification found 26 possible consistent deformations of scale-invariant rank-1 Seiberg-Witten geometries, of which 17 were found to correspond to known (i.e., constructed or predicted to exist by other methods) rank-1 SCFTs.  8 of the remaining 9 deformation geometries are found here as certain $\Z_n$-gauged versions of some of those 17 theories.  

In more detail, each entry in table \ref{tab1} describes a deformed rank-1 Seiberg-Witten geometry as $[K,F]$ where $K\in\{I_n, I_n^*, II, III, IV, II^*, III^*, IV^*\}$ is the Kodaira type of the scale-invariant singularity being deformed, and $F$ is the flavor symmetry which acts on the deformation parameters.\footnote{More precisely, only the Weyl group of the flavor symmetry acts on the deformation parameters.  Theories with flavor symmetries with the same Weyl group, such as $[II^*,G_2]$ and $[II^*, A_2 \rtimes \Z_2]$, have the same deformed Seiberg-Witten geometries \cite{allm1602}.}  In addition to the flavor symmetry, we will also denoted by $\chi_\d$ the existence of a chiral deformation parameter of scaling dimension $\d$; $\chi_0$ corresponds to the existence of an exactly marginal deformation.  If $F=\varnothing$ and there is no $\chi_\d$, then the corresponding SCFT has no relevant $\cN=2$ supersymmetry-preserving deformation.   The 17 known deformable theories referred to above are the entries in the ``parent" column of table \ref{tab1} excluding the ones in the light yellow rows which are free or IR-free theories, and excluding the non-deformable $[IV^*_{Q=1},\varnothing]$ geometry.   The 8 new geometries appear among the ones in the $\Z_2$ through $\Z_6$ columns of table \ref{tab1}, again excluding the free theories in the light yellow rows, and the two (more speculative) undeformable $[II^*,\varnothing]$ theories.

The 8 Coulomb branch geometries for which we find new SCFTs through discrete gauging are all characterized by the fact that upon deformation they flow to IR singularities --- such as $I_n^*$, $IV^*$, and $III^*$ --- which, by virtue of the scaling dimension of their Coulomb branch operator, or because of the Dirac quantization condition, cannot be consistently interpreted as corresponding to free theories.\footnote{For more on the analysis of such undeformable singularities see especially sections 1 and 4.2 of \cite{paper1}.}  We will show that these IR singularities can, in fact, all be identified as discretely gauged versions of IR-free $U(1)$ $\cN=2$ gauge theories.

We then argue that this identification can be extended consistently to interacting $\cN=2$ SCFTs.  That is, we realize the geometries that flow into these new IR singularities as the Coulomb branches of new SCFTs obtained by gauging discrete subgroups of other interacting $\cN=2$ SCFTs.  There are tight internal consistency checks stemming from the way the discrete symmetry being gauged acts on the Coulomb branch and on the deformation parameters, and from consistency under RG flows.  This allows only very special discrete symmetries to be gauged.  In particular, we will see that only certain combinations of $U(1)_R$ transformations, $SL(2,\Z)$ electric-magnetic (EM) duality transformations which act as discrete symmetries, and outer automorphisms of the flavor symmetry can be consistently gauged so as to preserve $\cN=2$ supersymmetry.  We indicate these in the $\Z_r$ columns in table \ref{tab1}.\footnote{Strictly speaking, in table \ref{tab1} we label the columns by the non-trivial group action of the gauged discrete symmetry on the CB, and not the gauged symmetry group itself.  In particular, for a given $\Z_r$ or $\green{\til \Z_r}$ in the table the actual gauged discrete symmetry is $\Z_{r\D}$ or $\green{\til\Z_{r\D}}$, where $\D$ is the scaling dimension of the CB of the parent theory; see the discussion in section \ref{setup}.}  For $\cN=4$ theories there is also a choice of gauging a discrete group preserving $\cN=3$ supersymmetry, indicated in table \ref{tab1} by the $\green{\til \Z_r}$ columns. This generalizes the identification by Garc\'ia-Etxebarria and Regalado in \cite{gr1512} of combinations of R-symmetry and $SL(2,\Z)$ transformations in $\cN=4$ theories that can be gauged so as to preserve $\cN=3$ supersymmetry. $\green{\til \Z_r}$ gauging gives different theories from the $\cN=2$ preserving $\Z_r$, yet since they both act in the same way on the CB (more below) and thus give the same daughter geometries, we do not distinguish them in our counting of the 26 consistent deformations.

In \cite{paper1} (see in particular the discussion at the end of section 4.2) we gave evidence that gauging discrete subgroups of the flavor symmetry does not preserve $\cN=2$ supersymmetry.  In this paper we will be able to present evidence that, likewise, gauging outer automorphisms of the flavor symmetry does not preserve $\cN=2$ supersymmetry unless combined as above with appropriate R-symmetry and $SL(2,\Z)$ transformations.  In free theories this can be seen explicitly, and arises from the way in which $U(1)$ gauge charges of local fields are correlated with their flavor charges.  For interacting theories where we have only a gauge-invariant description of the local fields this linkage forged by $\cN=2$ supersymmetry between the flavor symmetry and EM duality transformations is much less apparent.  In particular, it does not follow from properties of the (gauge invariant) local operator algebra of $\cN=2$ SCFTs, but instead must involve non-local (e.g., line) operators as well.  Nevertheless, we can infer this linkage in such theories from the structure of the effective action on the Coulomb branch (CB).  A classic example of this is the observation in \cite{sw2} that the $SL(2,\Z)$ EM-duality group of the $N_f=4$ $SU(2)$ gauge theory acts on the masses via outer automorphisms of the $\SO(8)$ flavor symmetry.  In a sense we generalize this observation to all rank-1 $\cN=2$ SCFTs, even those with no direct lagrangian description.

One, perhaps initially confusing, property of our results is that the same Coulomb branch geometry (i.e., the scale-invariant CB and its splitting under mass deformations \cite{paper1,paper2}) can occur for multiple distinct theories.  Physically, this just reflects the fact that two different microscopic theories can share the same low energy effective description.  For instance, in table \ref{tab1} the $[II^*,G_2]$ geometry appears as the Coulomb branch of a $\Z_3$-gauged ``daughter" of the $[I^*_0,D_4\Chi_0]$ theory (a.k.a., the $N_f=4$ $SU(2)$ gauge theory), and also as the Coulomb branch geometry of the $[II^*,A_2\rtimes\Z_2]$ SCFT.  Both are related, through RG flows, to the $[IV^*_{Q=1},\varnothing]$ singularity which appears as the Coulomb branch of either a $\Z_3$-gauged daughter of a free $\cN=2$ vector multiplet (in the $[II^*,G_2]$ case), or as the CB of an undeformable SCFT 
(as is in the case of $[II^*,A_2\rtimes\Z_2]$).  Another (more speculative) such example is the appearance of the $[II^*,\varnothing]$ singularity as two different undeformable SCFTs (in red in table \ref{tab1}) and as a $\Z_6$-gauged daughter of a free vector multiplet.  These and other examples in the table will be derived and explained in later sections.

For simplicity we will use, as above, the not strictly accurate terminology of parent and daughter theory, where the latter is a discretely gauged version of the former.  The inaccuracy arises for the reason briefly explained earlier: the parent and daughter theory have the same local dynamics and should really be considered as two different versions of the same theory and be treated on the same footing.  In terms of generalized global symmetries \cite{Gaiotto:2014kfa}, gauging a discrete $\Z_p$ global symmetry which acts on the local operators of the parent theory introduces a 2-form $\Z_p$ global symmetry under which surface operators of the daughter theory transform.  So, in principle, the existence of a discrete 2-form global symmetry is a way to know if a theory can be interpreted as a daughter of a parent theory without reference to the parent theory.  But we do not currently have control over the spectrum and symmetries of surface operators of the $\cN=2$ theories in question, and so will not pursue this approach here.\footnote{In class S much more is known about the spectrum of surface operators; see in particular \cite{Chacaltana:2012zy} for a discussion of discrete groups associated to such operators.}

The rest of the paper is organized as follows.  In section \ref{setup} we set up in generality the type of discrete symmetry group which can be gauged and in particular we derive the appropriate combination of $U(1)_R$ and $SL(2,\Z)$ transformations which preserve $\cN=2$ supersymmetry.  Section \ref{disgauge} is the heart of the paper where we systematically apply the results in section \ref{setup} to $\cN=2$ theories.  We start from a discussion of free $U(1)$ gauge theories and build our way up to discrete gauging of isolated non-lagrangian $\cN=2$ SCFTs. We also construct the $\cN=3$ preserving $\green{\til \Z_r}$ symmetries of $\cN=4$ theories.  In section \ref{sec:HB} we analyze the structure of the Higgs branches (HBs) of the discretely gauged SCFTs we constructed.  In particular we find that the HB of daughter theories is not the centered one-instanton moduli-space of the corresponding flavor group even if its parent theory's HB is.  The results we find are beautifully consistent with the constraints derived by the $\cN=2$ conformal bootstrap community \cite{b+13, Beem:2014rza, Lemos:2015orc, Liendo:2015ofa, Lemos:2015awa}.  In section \ref{sec:CC} we briefly discuss how the relation between the conformal central charges $a$, $c$ and $\D$, the scaling dimension of the CB vev, derived in \cite{st08} from the structure of the topologically twisted CB partition function is modified in discretely gauged SCFTs.  We finish by presenting our conclusions and a list of open questions.

\section{Discrete symmetries that preserve $\cN=2$ supersymmetry}\label{setup}

We start by reviewing the construction by Garc\'ia-Etxebarria and Regalado in \cite{gr1512} of an $\cN=4$ supersymmetric gauge theory with disconnected gauge group $O(2)\simeq \Z_2 \ltimes U(1)$.  This is a free $\cN=4$ $U(1)$ gauge theory (i.e., a free $\cN=4$ vector multiplet) with an additional gauged $\Z_2$ generated by
\begin{align}\label{N2act4}
C &: (A_\m, \l_\a^i, \vf^{[ij]}) \mapsto 
-(A_\m, \l_\a^i, \vf^{[ij]}), &
i,j &\in {\bf 4}\ \text{of}\ SO(6)_R,
\end{align}
where $(A_\m, \l_\a^i, \vf^{[ij]})$ are the propagating component fields of the vector multiplet.
This $\Z_2$ is just the charge conjugation symmetry of the $U(1)$ gauge theory, and commutes with the $\cN=4$ supercharges.  Since $C$ reverses the sign of $\vf^{[ij]}$, it quotients the moduli space by a $\Z_2$ action.  Thus the $\Z_2$-invariant moduli space parameters are the dimension-2 vevs of $(\vf^{[ij]})^2$. In \cite{gr1512} it was shown that the action of $C$ is equivalent to that of the element $(-I,-I) \in SO(6)_R \times SL(2,\Z)$, where $SO(6)_R$ is the R-symmetry group and $SL(2,\Z)$ is the discrete EM-duality group of of the $\cN=4$ theory.



From the $\cN=2$ perspective, the $U(1)$ $\cN=4$ is a theory of a free vectormultiplet and a massless neutral hypermultiplet. By giving a mass to the single neutral hypermultiplet and intergrating it out we obtain a free $U(1)$ $\cN=2$ gauge theory.  Following the construction above we can obtain an $O(2)$ $\cN=2$ gauge theory, by gauging the $\Z_2$ generated by 
\begin{align}\label{N2act}
C &: (A_\m, \l_\a^i, \f) \mapsto 
-(A_\m, \l_\a^i, \f), &
i &\in {\bf 2}\ \text{of}\ SU(2)_R.
\end{align}
In this case $C$ is a combination of $-I\in\SL(2,\Z)$ and $-1 \in U(1)_R$.   The flat CB of the $U(1)$ gauge theory is parameterized by $\f\in\C$ and has a trivial $SL(2,\Z)$ monodromy.  Quotienting by the action of $C$ gives a CB described by a flat $\C/\Z_2$ cone (i.e., with opening angle $\pi$) and an $\SL(2,\Z)$ monodromy of $-I$.

We extend these constructions to general $\cN=2$ theories as follows.  The continuous internal symmetries of an $\cN=2$ theory are the R- and flavor symmetries $U(1)_R \times SU(2)_R \times F$.  As argued in \cite{paper1} (section 4.2), discrete subgroups of $F$ cannot be gauged in an $\cN=2$ supersymmetry-preserving way without adding new degrees of freedom in the theory, while gauging a discrete subgroup of $SU(2)_R$ manifestly projects out some of the supercharges.  Thus only discrete $\Z_k \subset U(1)_R$ subgroups can be gauged.  

The theory may also have discrete global symmetries in $SL(2,\Z) \times \Out (F)$, where  $\Out (F)$ is the outer automorphism group of the flavor algebra.  So, we can consider gauging a discrete symmetry generated by a transformation 
\begin{align}\label{N2act1}
C=(\r,\s,\vf)\in U(1)_R\times SL(2,\Z)\times \Out (F).
\end{align}
These three factors affect the daughter theory in distinct ways:
\begin{itemize}
\item The action of the $\r\in U(1)_R$ factor on the CB vev, $u$, of the parent theory implies that upon gauging, the 
CB parameter of the daughter theory, $\til u$ will be given by $\til u=u^r$ with $r$ the smallest integer power necessary to build an operator invariant under the $U(1)_R$ action.  This thus increases the scaling dimension of the CB parameter of the daughter theory by a factor of $r$ relative to the parent theory.
\item The $\s\in SL(2,\Z)$ factor does not act on $u$ but fixes the value of the $U(1)$ gauge coupling, $\t$, of the daughter theory to particular values.  Note that only those subgroups of the $SL(2,\Z)$ EM duality group which fix $\t$ are global symmetries of the theory and can thus be gauged.  For example, a $\Z_4\in SL(2,\Z)$ can be gauged only for $\t=i$.
\item $\vf\in\Out (F)$ acts instead on the space of mass deformations, and thus disallows mass parameters which are not fixed by $\Out (F)$. The daughter theory will then have a flavor symmetry algebra $F':=F/\Out (F)$.
\end{itemize}
Here we are only interested in combinations of these factors which preserve (at least) $\cN=2$ supersymmetry.  First recall that under an $SL(2,\Z)$ transformation, $\s :=\left(\begin{smallmatrix}a&b\\ c&d\end{smallmatrix}\right)$, the $U(1)$ coupling transforms as $\s:\ \t\to\frac{a\t+b}{c\t+d}$, and the chiral supercharges transform by a phase \cite{Kapustin:2006pk}
\beq\label{SLtr}
\s:\  Q_\a^i
\to
\left(\frac{|c\t+d|}{c\t +d}\right)^{1/2}
Q_\a^i .
\eeq

\begin{table}
\centering
$\begin{array}{|c|l|c|c|c|c|c|}
\hline
\multicolumn{7}{|c|}{\text{\bf Possible scaling behaviors near singularities of a rank 1 CB}}\\
\hline\hline
\text{Name} & \multicolumn{1}{c|}{\text{regular SW curve}} & \ \text{ord}_0(D_{x})\ \ &\ \D(u)\ \ & M_{0} & \text{deficit angle} 
& \t_0 \\
\hline
II^*   &\parbox[b][0.45cm]{4cm}{$\ y^2=x^3+u^5$}             
&10 &6 &ST &\pi/3 & \ e^{i\pi/3}\ \\
III^*  &\ y^2=x^3+u^3x &9 &4 &S &\pi/2 & i\\
IV^*  &\ y^2=x^3+u^4 &8 &3 &-(ST)^{-1} &2\pi/3 & e^{2i\pi/3}\\
I_0^* &\ y^2=\prod_{i=1}^3\left(x-e_i(\t)\, u\right)
&6 &2 &-I &\pi & \t\\
IV &\ y^2=x^3+u^2 &4 &3/2 &-ST &4\pi/3 & e^{2i\pi/3}\\
III &\ y^2=x^3+u x &3 &4/3 &S^{-1} &3\pi/2 & i\\
II  &\ y^2=x^3+u &2 &6/5 &(ST)^{-1} &5\pi/3 &e^{i\pi/3}\\
I_0  &\ y^2=\prod_{i=1}^3\left(x-e_i(\t)\,\right)
&0 &1 &I &0 & \t\\
\hline
\hline
I^*_n\ \ (n{>}0) &
\parbox[b][0.45cm]{5cm}{
$\ y^2=x^3+ux^2+\L^{-2n}u^{n+3}\ \ $}
& n+6 & 2 & {-T^n} & 2\pi\ \text{(cusp)} 
& i\infty\\
I_n\ \ (n{>}0)    &\ y^2=(x-1)(x^2+\L^{-n}u^n)  
& n     & 1 & {T^n} & 2\pi\ \text{(cusp)} 
& i\infty\\[0.5mm]
\hline
\end{array}$
\caption{\label{Table:Kodaira} Scaling forms of rank 1 special K\"ahler singularities, labeled by their Kodaira type (column 1), a representative family of elliptic curves with singularity at $u=0$ (column 2), order of vanishing of the discriminant of the curve at $u=0$ (column 3), mass dimension of $u$ (column 4), a representative of the $SL(2,\Z)$ conjugacy class of the monodromy around $u=0$ (column 5), the deficit angle of the associated conical geometry (column 6), and the value of the low energy $U(1)$ coupling at the singularity (column 7).  The first eight rows are scale invariant.  The $I_0$ ``singularity" in the eighth row is the regular (flat) geometry corresponding to a free vector multiplet.  The last two rows give infinite series of singularities which have a further dimensionful parameter $\L$ so are not scale invariant; they are IR free since $\t_0=i\infty$.}
\end{table}

Now, a $\Z_k\subset SL(2,\Z)$ is only a symmetry of the theory for values of $\t$ fixed by the $\Z_k$ action.  The possible scale invariant CB geometries have fixed values of $\t$, and therefore fix the subgroup of $SL(2,\Z)$ which acts as a symmetry.  Table \ref{Table:Kodaira} lists the possible scaling behaviors of singularities on a rank-1 CB and their properties.  (Their naming follows Kodaira's for degenerate fibers of elliptic surfaces \cite{KodairaI, KodairaII}.)   It follows that the subgroup of $SL(2,\Z)$ which is a symmetry for a given CB geometry is
\begin{align}\label{allowedsigma}
\begin{array}{l|c|c}
\text{singularity} &\ \text{subgroup of $SL(2,\Z)$}\ 
&\text{generators}\\
\hline
\text{$II$, $II^*$, $I_0^*$, or $I_0$ at $\t=e^{i\pi/3}$} 
& \Z_6 & \s_6 = ST \\
\text{$III$, $III^*$, $I_0^*$, or $I_0$ at $\t=i$} 
& \Z_4 & \s_4 = S\\
\text{$IV$, $IV^*$, $I_0^*$, or $I_0$ at $\t=e^{2i\pi/3}$}\ 
& \Z_2 \times \Z_3 &\ \s_2=-I, \ \s_3=-ST\ \\
\text{$I_0^*$ or $I_0$ at any other $\t$} 
& \Z_2 & \s_2 = -I
\end{array}
\end{align}
where the $S$ and $T$ generators of $SL(2,\Z)$ are $T=\left( \begin{smallmatrix} 1&1\\ 0&1 \end{smallmatrix} \right)$, $S=\left( \begin{smallmatrix} 0&-1\\ 1&0 \end{smallmatrix} \right)$.  The $\s$ generators listed in \eqref{allowedsigma} are just representatives of their $SL(2,\Z)$ conjugacy class, and also could equally well be replaced by their inverses.  Furthermore, since $\Z_6 \simeq \Z_2 \times \Z_3$, the first and third lines in \eqref{allowedsigma} have the same discrete symmetry.  Indeed, since $S^2=(ST)^3=-I$, the $\Z_2$ subgroup in each case is the center of $SL(2,\Z)$, generated by $\s_2=-I$.  It is then easily checked that for any $\s$ generating a $\Z_k$ subgroup of $SL(2,\Z)$ with the associated value of $\t$ shown in \eqref{allowedsigma}, \eqref{SLtr} reduces to 
\begin{align}\label{SLtr1}
SL(2,\Z) \supset\Z_k \ni \s:\  Q_\a^i \to {\rm e}^{-i\pi/k}Q_\a^i.
\end{align}

The chiral supercharges, in the normalization of \cite{paper1}, have $U(1)_R$ charge $R(Q_\a^i)=\D(Q_\a^i)=1/2$.  It then follows that by choosing $\r$ to be the generator of a $\Z_k\subset U(1)_R$, the $(\r,\s)\in U(1)_R\times SL(2,\Z)$ transformation generates a $\Z_k$ group which leaves both supercharges invariant, and thus preserves $\cN=2$ supersymmetry.

In general the $\Z_k\subset U(1)_R$ generated by $\r$ acts non-trivially as a $\Z_r$ on the CB of the parent theory, where 
\begin{align}\label{rdef}
r := \frac{k}{\ell},\qquad
\text{with}\quad \ell := \D(u),
\end{align}
and where $\D(u)$ is the scaling dimension of the CB parameter of the parent theory.  This is readily seen by noticing that the CB parameter $u$ is identified as the vev of the conformal primary of the $\cE_{\ell\,(0,0)}$ superconformal multiplet, which is a scalar $SU(2)_R$ singlet with $U(1)_R$ charge $\ell=\D(u)$ \cite{paper1}.  Thus under the $\Z_k\subset U(1)_R$ the CB parameter transforms as
\beq\label{eqU1R}
\r:\ u \to {\rm e}^{i2\pi\ell/k} u.
\eeq
It then follows that upon gauging this symmetry, the CB of the daughter is parametrized by $\til u=u^{k/\ell}$ so that $\D(\til u)=k$.   Since $r$ is integer, it follows that $\ell=\D(u)$ should divide $k$.\footnote{The $k$ and $\ell$ defined here are the same those in \cite{Aharony:2016kai}.}  


We will see in the next section how the choice of $\vf\in\Out (F)$ is tied to $\r$ and $\s$.

\section{Discrete gauging of $\cN=2$ theories}\label{disgauge}
   
Having established the general structure of the possible discrete symmetries which preserve $\cN=2$ SUSY we can now systematically build our way up to discrete gauging of non-lagrangian $\cN=2$ SCFTs. We will first present a discussion of $O(2)$ theories, that is $\Z_2$ gauging of $U(1)$ gauge theories with matter, and of $\Z_k$ gaugings for $k\neq 2$ of free $U(1)$ theories.\footnote{We do not consider $\Z_k$ gaugings with $k\neq2$ of $U(1)$ theories with matter, and, in fact, will show below that such $\Z_k$ gaugings are inconsistent with $\cN=2$ supersymmetry.}  This generalization is the starting point for the following analysis of non-lagrangian theories and then of $\cN=3$ theories.  In particular, the CB analysis of $O(2)$ theories with matter is one of the crucial tools which we will use to show the consistency under mass deformation of the discretely gauged non-lagrangian theories which we will construct. 

\subsection{$O(2)$ theories with matter}\label{U1matt}

In this section we want to extend the gauging of \eqref{N2act1} to IR-free $U(1)$ gauge theories with general matter content.  We start with the case where $C=(\r,\s,\vf)$ generates a $\Z_2$ with action on the vector multiplet as in \eqref{N2act}.

First of all notice that a $\Z_2$ gauging cannot always be implemented since, for a non-scale invariant theory, the $U(1)_R$ symmetry (in which the $\r$ factor of the $\Z_2$ generator is embedded) is anomalous.  In particular, it is broken down to $\Z_n$, where $n = \sum_I q_I^2$ is the coefficient of the one-loop beta function and $q_I$ is the $U(1)$ gauge charge of the $I$-th hypermultiplet.  If $n$ is odd $\Z_2\not\subset\Z_n$, implying that $\r$ in \eqref{N2act1} is anomalous.  

This obstruction can also be seen from the CB geometry: for odd $n$ no consistent CB geometry for the discretely gauged theory can be constructed.  Carefully analyzing how this obstruction arises from the CB geometry is a good warm-up for the studies which follow where the CB data will be the only information available. 

To discuss the CB geometry after the $\Z_2$ gauging, it is helpful to recall that the CB geometry of an $\cN=2$ $U(1)$ gauge theory with massless matter only depends on the value, $n$, of its beta function coefficient.  It is the $I_n$ geometry in table \ref{Table:Kodaira} which has a single\footnote{Here we are only discussing the region around the origin of the moduli space, neglecting the $|u|\gtrsim \L$ region, where $\L$ is the Landau pole scale.} cusp-like singularity located at $u=0$, with a $T^n$ monodromy.  The $\Z_2$ transformation \eqref{N2act} acts as a $\pi$ rotation on the CB: $u\mapsto -u$. The fact that the $\Z_2$ in \eqref{N2act} is a symmetry, translates into the fact that the values of the special coordinates at $u$ and $-u$ are equal up to an $SL(2,\Z)$ transformation $M$.  Performing \eqref{N2act} twice corresponds to going around the $u=0$ singularity by a full $2\pi$ and thus $[M^2]=[T^n]$.  (Square brackets indicate $SL(2,\Z)$ conjugacy classes.)  This constraint cannot always be satisfied.  It is easy to show that such $M$ only exists for even values of $n$, and only two solutions, $[M]=[\pm T^{n/2}]$, are allowed up to conjugation.  By construction $M$ will be the ``effective'' monodromy of the CB geometry of the daughter $O(2)$ theory.  The resultant CB geometry is parametrized by $\til u := u^2$.  From table \ref{Table:Kodaira} we can see that only $[M]=[-T^{n/2}]$ is compatible with a scaling dimension 2 CB parameter.  Thus, after gauging a $\Z_2$, a parent $I_{2n}$ CB becomes a daughter $I_n^*$ CB. This can also be seen explicitly working with the $I_{2n}$ curve. Carefully performing the discrete gauging we obtain the curve describing the $I^*_n$ singularity, see appendix \ref{A1} for the explicit calculation.

This picture is not the whole story, as discrete gauging also relates the deformations of the parent and daughter theories.  Recall that mass deformations of both the parent and daughter theory appear as vevs of vector multiplets upon weakly gauging their flavor symmetries, $F$ and $F'$ respectively.  So their mass parameters  can be thought of as linear coordinates on the complexified Cartan subalgebras of $F$ and $F'$.  Thus to discuss the allowed mass deformations of the daughter theory we need to understand how \eqref{N2act} acts on the flavor symmetry algebra of the parent theory.

Let us start by studying the simplest non-free $\cN=2$ theory: a $U(1)$ gauge theory with a single charge 1 hypermultiplet. We can express the hypermultiplet as a doublet $(Q_+,Q_-)$ of $\cN=1$ chiral superfields with charges $\pm1$.  The theory has a $U(1)_F$ flavor symmetry and a single mass deformation.  $\cN=2$ SUSY implies the existence of a term in the lagrangian (written in $\cN=1$ superfield language)
\beq\label{N2coupling}
\sim \int{\rm d}^2\th\ Q_+\Phi\ Q_- 
\eeq
where $\Phi$ is the $\cN=1$ chiral superfield in the $\cN=2$ vector supermultiplet.  Consider now gauging a $\Z_2$ symmetry which acts on the vector multiplet as in \eqref{N2act}, and so as $\Phi\mapsto-\Phi$.  To preserve $\cN=2$ SUSY, \eqref{N2coupling} needs to be invariant which implies that $Q_+Q_-$ must pick up a sign under the $\Z_2$.  This in turns dictates the transformation of the $\cN=2$ mass term:
\beq\label{N2mass}
C:\quad \int{\rm d}^2\th\ m\ Q_+Q_-
\quad\mapsto\quad
-\int{\rm d}^2\th\ m\ Q_+Q_- .
\eeq
\eqref{N2mass} can be reinterpreted as an action of $\Out (U(1)_F)$ on the mass parameter.   Indeed, $\Out (U(1)_F) =\Z_2$ is generated by the complex conjugation automorphism of $U(1)_F$ which acts on the Cartan subalgebra by a reflection through the origin, $m \mapsto -m$.

This calculation shows that gauging a discrete subgroup of $U(1)_R$ and $SL(2,\Z)$ in an $\cN=2$ supersymmetry-preserving way requires the discrete group to also have an $\Out (F)$ action as in \eqref{N2act1}.  In the general case where $F$ is not just $U(1)_F$, but may have many $U(1)$ and simple factors, a more subtle argument is needed to identify which $\vf\in\Out (F)$ needs to be discretely gauged.  The key point of the flavor $U(1)_F$ example in the last paragraph was that the flavor and gauge charges of the (gauge-variant) local fields $Q_\pm$ appearing in the lagrangian are correlated.  The charge conjugation symmetry of \eqref{N2act} implies that it must interchange $Q_+$ with $Q_-$.  When combined with $\cN=2$ supersymmetry, which we showed in the last paragraph implies $Q_+Q_- \mapsto -Q_+Q_-$, this implies that we must choose $C$ to act on the hypermultiplets as $C:\ Q_\pm \mapsto \pm e^{\pm i\a} Q_\mp$.  The $e^{\pm i\a}$ factors are just an arbitrary $U(1)_F$ flavor phase rotation, which can be removed by flavor rotating $Q_\pm \to e^{\mp i\a/2} Q_\mp$, so that we can put $C$ into a canonical form $C:\ Q_\pm \mapsto \pm Q_\mp$.  (Note that $C^2 = -1$, so that $C$ actually generates a $\Z_4$ action on the gauge-variant local fields, though it only acts non-trivially as a $\Z_2$ on gauge-invariant combinations of local fields.)

We will now generalize this to the case where there are $n$ massless hypermultiplets all with $U(1)$ gauge charge $\pm1$.  This theory has $F=U(n)\simeq U(1)\times SU(n) \simeq U_1A_{n-1}$ flavor symmetry\footnote{We will often use Dynkin notation for simple Lie algebras together with ``$U_1$" to stand for $U(1)$ factors.  Thus $U(3) \simeq U(1)\times SU(3) \simeq U_1A_2$.  Also, it will be useful to keep in mind the low-rank degeneracies of the Dynkin notation: $A_1=B_1=C_1$, $D_1=U_1$, $B_2=C_2$, $D_2=A_1A_1$, and $D_3=A_3$.} under which $Q^i_+$ and $Q_{-i}$, $i=1,\ldots,n$, transform in the ${\bf n}_+$ and $\bar{\bf n}_-$ flavor representations, respectively.  Then \eqref{N2coupling} reads $\int d^2\th\, Q^i_+\Phi Q_{-i}$, so invariance under the $\Z_2$ requires only that $C:\, Q^i_+Q_{-i} \mapsto - Q^i_+Q_{-i}$, while the charge conjugation action of $C$ requires that $Q^i_+$ and $Q_{-i}$ be interchanged.  The general solution for the linear action of $C$ on the local fields is
\begin{align}\label{}
C_f:\ 
\begin{cases}
Q^i_+ &\mapsto\  + f^{ij}\, Q_{-j}\\
Q_{-i} &\mapsto\  - Q^j_+\, (f^{-1})_{ji}
\end{cases} , \qquad f\in U(n).
\end{align}
Here we have labelled the $C$ action by the choice of element $f$ of the flavor group.
Since the $\cN=2$ mass term is $\int d^\th\, Q^i_+ {m_i}^j Q_{-j}$, the action of $C_f$ on the flavor adjoint masses is
\begin{align}\label{OutFact}
C_f:\ m \mapsto -f^{-1} m^T f,
\end{align}
in the obvious matrix notation.  Since $m$ is an element of the Lie algebra of $F$, this gives action on $\ff:=\text{Lie}(F)$ which is easily checked to be an automorphism of $\ff$.  It generates a subgroup $\til\G_f\subset\Aut(\ff)$ which lifts to a subgroup $\G_f\subset\Out(\ff) \simeq \Aut(\ff)/\Inn(\ff)$ where $\Inn(\ff)$ is the group of inner automorphisms which are automorphisms whose actions on $\ff$ are all of the form $m\mapsto f^{-1} m f$ with $f\in F$.  Thus the flavor symmetry algebra, $\ff'$, of the daughter theory is $\ff' =\ff/\G_f$, generated by those elements of $\ff$ left invariant by \eqref{OutFact}.  

In \eqref{OutFact} $f$ is undetermined. Note, however, that the daughter flavor symmetry, $\ff'$, can depend on the specific choice of $f$ defining $C_f$ in \eqref{OutFact}.  Even though, as an abstract group, $\G_f \subset \Out(\ff)$ generated by $C_f$ is independent of $f$, its action on $\ff$ is not.\footnote{We thank Y. Tachikawa for explaining this to us.}  A set of rules governing what are the inequivalent $\ff'$ that can result from different choices of $f$ are summarized in section 3.3 of \cite{Tachikawa:2011ch}.  (See \cite{fuchs2003symmetries, Kac:1994IDL} for more detailed discussions of automorphisms of Lie algebras.)

We do not know what determines the choice of $f$ in \eqref{OutFact}.  Nevertheless, we do know that not all such choices are consistent with $\cN=2$ supersymmetry. This follows from demanding a consistent action of the $\Z_2$ discrete symmetry on the CB geometry of the theory, which we will discuss shortly.  For instance, the choice of $f=$id$\in F$ might seem ``natural", however, with this choice $\ff' = D_{n/2}$ for $n$  even, while the CB analysis implies that Weyl$(\ff')$ is of $BC_{n/2}$ type (see appendix \ref{appA}). Furthermore there is always a choice of $f\equiv \til f$ for which the $\ff'$ is obtained as folding of the Dynkin diagram of the flavor symmetry algebra of the parent theory. From our analysis this choice seems always compatible. When we will talk about the Out($F$) action in what follows below, unless otherwise stated, we will implicitly assume $f=\til f$. Perhaps our inability of determining the right element $f$ is related to the puzzle of discretely gauging subgroups of $\Inn(F)$ in a way consistent with $\cN=2$ supersymmetry, pointed out at the end of section 4.2 of \cite{paper1}.

For general hypermultiplet content $\{ Q_{\pm I}\}$ consisting of $n_I$ hypermultiplets with $U(1)$ gauge charge $\pm I$ for some set of charges $\{I\}$, the flavor symmetry is $F=\prod_{I} U(n_I)$.   Its outer automorphism group is $\Out (F) = \prod_I (\Z_2\times\Z_2)$, since each $U(n_I)\simeq U(1)\times SU(n_I)$ factor contributes a $\Z_2$ from the $U(1)$ complex conjugation and another $\Z_2$ from the $SU(n_I)$ complex conjugation automorphisms.  Now, for a given charge $I$, the $n_I$ $Q_{+I}$ and $Q_{-I}$ fields transform in the $({\textbf n_I})_{+1}$ and $(\overline{\textbf{n}}_I)_{-1}$, respectively, of the $U(n_I)$ flavor factor.  Thus, charge conjugation, which reverses the $U(1)$ gauge charges of all fields, will necessarily also complex conjugate all their flavor charges.  Thus it is the overall ``diagonal" $\Z_2^\text{diag} \subset \prod_I (\Z_2\times\Z_2) \simeq \Out (F)$ which is generated by $\vf\in\Out (F)$ appearing in \eqref{N2act1}.

Only the mass deformations which are invariant under this $\Z_2^\text{diag}$ survive as mass deformations of the daughter $O(2)$ gauge theory.  Thus the flavor symmetry algebra of the daughter theory will be $\ff' = \oplus_I A_{n_I-1}/(\Z_2)_I$ where $(\Z_2)_I\subset \Out(A_{n_I-1})$ acts as in \eqref{OutFact} for some choice of $f_I \in SU(n_I)$.  The evidence from demanding a consistent action on the CB geometry is that these $f_I$ must be chosen so that $\ff' = \oplus_I BC_{[n_I/2]}$ where the square brackets mean geratest integer part and the $BC_n$ notation just reflects our inability to distinguish between the $B_n$ and $C_n$ possibilities on this basis.

\begin{center}
---\ \ $*$\ \ ---
\end{center}

We now describe how this $\Z_2$ gauging is reflected in the CB geometry of the parent and daughter theories.  Since all the objects appearing in the low energy effective action on the CB are gauge invariant, the way the above correlation of gauge and flavor charges in the microscopic gauge theory description appears in the CB geometry is indirect.

For simplicity and concreteness, we will illustrate this with a $U(1)$ gauge theory with $3$ hypermultiplets of charge $\pm\sqrt{2}$.\footnote{This somewhat unusual choice of charge assignment is due to the facts that (i) the ambiguity on what subgroup of the outer autormorphism group needs to be gauged only arises with three or more hypers, and (ii) the $U(1)$ gauge theory with $3$ hypers of charge $1$ has odd beta function for which no $\Z_2$ discrete gauging is allowed.} This theory has a $U(3)\cong U_1 A_2$ flavor symmetry and an $I_6$ CB geometry, which under discrete gauging is transformed, following the discussion above, into an $I_3^*$ CB geometry.  The flavor outer automorphism group is $\Out (U(3))= \Out (U_1)\times \Out (A_2) =\Z_2^{(1)} \times \Z_2^{(2)}$, where we denote $\Z_2^{(1)} = \Out (U_1)$ and $\Z_2^{(2)} = \Out (A_2)$.  

Now introduce the gauge-invariant ``meson" (or moment map) operators $M^j_i := Q^j_+ Q_{i-}$, $i,j=1,2,3$.  $\cN=2$ supersymmetry implies the superpotential term in the action of the form $\int d^2\th\ \Phi \sum_{j=1}^3 M^j_j$ as in \eqref{N2coupling}, so invariance under \eqref{N2act} implies only that the meson operator satisfies
\beq\label{N2couplingNf}
C:\quad \sum_{j=1}^3 M_j^j
\quad\mapsto\quad
-\sum_{j=1}^3 M_j^j ,
\eeq
which does not determine a unique action of $C$ on the local gauge-invariant operators $M^i_j$.  Since a general mass deformation can be written (up to a flavor transformation) as
\beq\label{N2massNf}
\sum_{j=1}^3 m_j M^j_j ,
\eeq
it also follows that \eqref{N2couplingNf} does not dictate a unique action of $\Out (U_1A_2)$ on the masses: both $\Z_2^\text{diag}\subset\Z^{(1)}_2\times\Z^{(2)}_2$ as well as the $\Z^{(1)}_2\subset\Z^{(1)}_2\times\Z^{(2)}_2$ are compatible with \eqref{N2couplingNf}.  It would thus appear that we could construct two different $I_3^*$ CB geometries, one with flavor group\footnote{As explained above, although the subgroup of the outer automorphism subgroup which participates in $\Z_2$ action is uniquely determined, the identification of the daughter flavor group is not.  The $BC_n$ notation reflects this ambiguity: from the CB geometry the Weyl group of the daughter flavor symmetry is of BC type, so the (maximal) daughter flavor algebra is either $B_n$ or $C_n$.  (Of course, for $n=1,2$, these two algebras happen to be isomorphic.)} $(U_1A_2)/\Z_2^\text{diag}\cong BC_1$ and one with $(U_1A_2)/\Z^{(1)}_2\cong A_2$.  But we have seen above from the lagrangian description that only the former is allowed, and we will now explain why it is the only one which gives a consistent CB geometry under deformation.

For a generic mass deformation with masses $(m_1,m_2,m_3)$ as in \eqref{N2massNf} the $I_6$ singularity splits into three separate $I_2$ singularities \cite{paper1} at $u=m_j$, $j=1,2,3$, each one associated with a single hypermultiplet of charge $\sqrt{2}$ becoming massless.  It is easy to see that the generic mass deformations invariant under the two choices, $\Z_2^\text{diag}$ and $\Z_2^{(1)}$, of the outer automorphism group action are (up to the action of the Weyl group of $U_1A_2$)
\begin{align}\label{}
(a)\quad \Z_2^\text{diag} &\longleftrightarrow
\left\{\begin{array}{l}
m_1\to\mu\\
m_2\to0\\
m_3\to-\mu
\end{array}\right.,
&(b)\quad \Z^{(1)}_2 &\longleftrightarrow
\left\{\begin{array}{l}
m_1\to\mu\\
m_2\to\nu\\
m_3\to-\mu-\nu
\end{array}\right..
\end{align}
The arrangement on the CB of the of the three $I_2$ singularities under the deformations $(a)$ and $(b)$ is depicted in figure \ref{FigSpl}.  (We give the explicit SW curve describing the maximally deformed $I_n$ CB geometry in appendix \ref{A1}.)  It is evident that only mass deformation $(a)$ gives a CB geometry which can be consistently quotiented by the $\Z_2$ action in \eqref{N2act} which, as we described earlier, acts by $\r: u\mapsto -u$ on the CB.  We thus conclude that the only $\Z_2$ symmetry whose gauging is allowed by $\cN=2$ supersymmetry gives rise to an $I_3^*$ with a $BC_1\cong A_1$ flavor symmetry algebra, which we denote as the $[I_3^*,BC_1]$ theory.

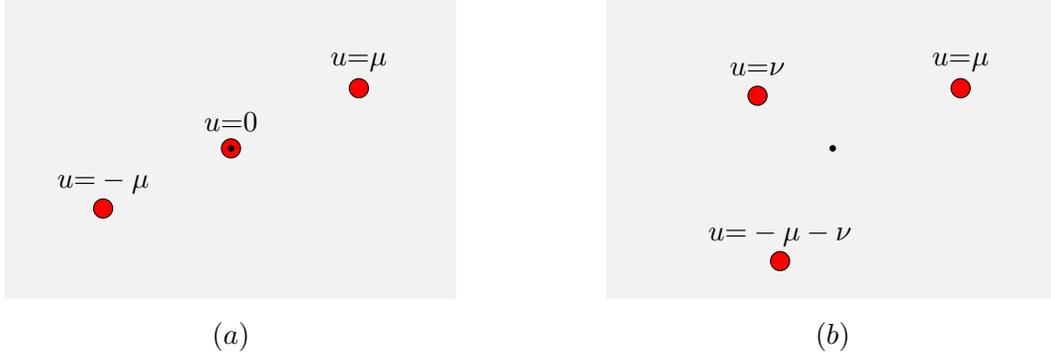
\begin{figure}[tbp]
\centering
\begin{tikzpicture}
\begin{scope}[xshift=0cm]
\fill[color=black!05] (0,0) rectangle (6,4);
\node[R] at (3,2) {};
\node[bl] at (3,2) {};
\node[R] at (4.7,2.8) {};
\node[R] at (1.3,1.2) {};
\node at (3,2.35) {$u{=}0$};
\node at (4.7,3.15) {$u{=}\m$};
\node at (1.3,1.55) {$u{=}-\m$};
\node at (3,-.5) {$(a)$};
\end{scope}
\begin{scope}[xshift=8cm]
\fill[color=black!05] (0,0) rectangle (6,4);
\node[bl] at (3,2) {};
\node[R] at (3-1,2+.7) {};
\node[R] at (4.7,2.8) {};
\node[R] at (1.3+1,1.2-.7) {};
\node at (3-1,2.35+.7) {$u{=}\n$};
\node at (4.7,3.15) {$u{=}\m$};
\node at (1.3+1,1.55-.7) {$u{=}-\m-\n$};
\node at (3,-.5) {$(b)$};
\end{scope}
\end{tikzpicture}
\caption{Singularities on a deformed $I_6$ CB for two different mass deformations.  The red circles mark the positions of the $I_2$ singularities, and the black dot marks the origin.}
\label{FigSpl}
\end{figure}

Let's now explicitly perform the quotient of the deformed $(a)$ geometry and see that it is in fact consistent.  Under the $\Z_2$ action on the CB the two $I_2$ singularities located at $u=\pm \mu$ are identified while for the $I_2$ singularity at the origin we can apply the reasoning from the beginning of this section to conclude that it becomes an $I_1^*$ singularity.  Thus the initial $I_3^*$ singularity of the daughter theory splits under the mass deformation allowed by the discrete gauging into an $I_2$ singularity at $\til u = u^2 =\mu^2$ and an $I_1^*$ singularity at $\til u =0$.  This is summarized by the commutative diagram
\beq\label{defdau}
\begin{array}{ccc}
I_6&\xrightarrow{(U_1A_2)/\Z^\text{diag}_2\ \text{deformation}}
&\{{I_2}^3\}\\
\quad\Big\downarrow{\Z_2}&& \quad\Big\downarrow{\Z_2}\\
I^*_3&\xrightarrow{\ \ \ \ B_1\ \text{deformation}\ \ \ \ }
&\{I_1^*,I_2\}
\end{array} .
\eeq
This quotient of the CB geometry is demonstrated explicitly as an operation of the SW curve in appendix \ref{appA}.

As a further check, one of the conditions for a deformation of a CB singularity to be consistent is that the sum of the orders of vanishing of the Seiberg-Witten curve discriminants at the singularities after the splitting is an invariant of the deformation \cite{paper1}.  We can read off the orders of the singularities involved in the splitting in \eqref{defdau} from table \ref{Table:Kodaira}, to find, consistently, that both $I^*_3$ and $\{I^*_1,I_2\}$ have order 9.  In fact, this condition is enough to select \eqref{defdau} as the only possible consistent deformation pattern:  because of the nature of its parent theory and the $\Z_2$ gauging, a generic deformation pattern for a discretely gauged $I_3^*$ can only be of the form $\{{I_2}^k, I_n^*\}$, where the $I_n^*$ arises from the action of the $\Z_2$ on the origin of the CB.  It is straightforward to see that only $(k=0,n=3)$ and $(k=1,n=1)$ give a consistent deformation pattern. The latter is what we just described while the former would be equivalent to a frozen $I_3^*$.  But $\Out(U(3))$ is not large enough to entirely freeze all mass parameters of the parent theory, so we are thus led to discard the second option, as well as concluding that upon gauging the $\Z_2$, $[I_6, U_1A_2]\mapsto [I_3^*, BC_1]$.

\begin{table}[h!]
\small\centering
\centering
$\begin{array}{c|c|ccc|c}
\multicolumn{6}{c}{\ccy}\\
\multicolumn{6}{c}{\multirow{-2}{*}{\text{\large CB geometries of some $\cN=2$ supersymmetric $O(2)$ gauge theories\ccy}}}\\
\hhline{======}
\text{CB geometry}  & \text{field theory content} & \text{flavor} 
& & \text{CB geometry} & \text{flavor} \\
\hhline{===~==}
I_0 &\text{free vector multiplet} & \varnothing & 
\multirow{11}{*}{ \xdasharrow{\text{After $\Z_2$ gauging\ \ }}} & I_0^* & \varnothing \\
\hhline{---~--}
\multirow{2}{*}{$I_2$}&\text{w/ 2 $Q{=}1$ hypers} & U(2) &  &\multirow{2}{*}{$I_1^*$}& BC_1 \\
&\quad \text{w/ 1 $Q{=}\sqrt{2}$ hyper} & U(1) &  & &\varnothing \\
\hhline{---~--}
\multirow{3}{*}{$I_4$}& \text{w/ 4 $Q{=}1$ hypers} & U(4) &  &\multirow{3}{*}{$I_2^*$}& BC_2 \\
&\text{w/ 2 $Q{=}\sqrt{2}$ hypers} & U(2) &  & &BC_1\\
&\text{w/ 1 $Q{=}2$ hyper} & U(1) &  & &\varnothing\\
\hhline{---~--}
\multirow{5}{*}{$I_6$}& \text{w/ 6 $Q{=}1$ hypers} & U(6) &  &\multirow{5}{*}{$I_3^*$}& BC_3 \\
&\text{w/ 3 $Q{=}\sqrt{2}$ hypers} & U(3) &  & & BC_1 \\
&\text{w/ 2 $Q{=}\sqrt{3}$ hypers} & U(2) &  & & BC_1 \\
&\text{\ w/ 1 $Q{=}2$ and 2 $Q{=}1$ hypers\ } &\ U(1){\times} U(2)\ & & & BC_1 \\  
&\text{w/ 1 $Q{=}\sqrt6$ hyper} & U(1) & & & \varnothing\\
\hhline{======}
\end{array}$
\caption{\label{Geometries}
All consistent $\Z_2$ discrete gaugings of the $I_{0,2,4,6}$ singularities.  The three leftmost columns show the CB geometries, field content, and flavor symmetries of the ``parent" $U(1)$ gauge theories.  The two rightmost columns show the CB geometries and flavor symmetries of the resulting ``daughter" $O(2)$ gauge theories after discretely gauging the appropriate $\Z_2$ symmetry.}
\end{table}

This same reasoning can easily be extended to gauging $\Z_2$ global symmetries in other $U(1)$ gauge theories in an $\cN=2$ supersymmetry-preserving way.  Indeed, it is both instructive and useful for further reference to explicitly carry out all possible $\Z_2$ gaugings of $I_{2n}$ singularities for small values of $n$.  The results are reported in table \ref{Geometries}.    

Summarizing: 
\begin{itemize}
\item A $\Z_2$ gauging of an $\cN=2$ theory with $I_m$ CB and flavor symmetry algebra $F$ can only be done if $m=2n$ is even.  If $m$ is odd the $\Z_2$ is anomalous.
\item The daughter theory has an $I_n^*$ CB geometry described by a CB parameter $\til u$ with scaling dimension $\D(\til u)=2$.
\item The daughter theory has flavor symmetry algebra $F'=F/\Z_2^\text{diag}$ where $\Z_2^\text{diag}\subset\Out(F)$ is the ``diagonal'' flavor outer automorphism subgroup which acts on both the $U(1)$ and non-Abelian factors of the flavor symmetry algebra $F$.
\end{itemize}

$I_n^*$ singularities also arise as the CB geometries of $\cN=2$ IR-free $SU(2)$ gauge theories with beta function equal to $n$ \cite{paper1, paper2}.  The ones constructed through discrete gauging can be distinguished from the ones arising in $SU(2)$ gauge theories by their different flavor groups and spectrum of mass operators.  
In particular, discrete gauging allows the construction of  ``frozen'' versions of $I_n^*$ singularities for any $n$; see, e.g., table \ref{Geometries}.
These ``frozen'' $I_n^*$ will play an important role in later sections since they will arise as IR fixed points of RG flows from non-lagrangian discretely gauged theories.

Even though the spectrum of the local gauge invariant operators in the microscopic theory (i.e., at the scale-invariant vacuum sector) does not provide enough information to infer the action of discrete gauging on the flavor symmetry, the low energy effective theory as encoded in the geometry of the CB does provide the needed information.  In the lagrangian theories so far discussed we had access to the local gauge-variant fields which carry both gauge charges and flavor charges. We could thus determine the required subgroup of flavor outer automorphisms which must accompany the charge conjugation $\Z_2$.  In strongly-coupled non-lagrangian theories where we do not have a description in terms of gauge-variant local fields, and in cases where we will not be gauging a charge conjugation $\Z_2$ (but instead some other discrete symmetry which acts non-trivially on the gauge sector via its embedding in the $SL(2,\Z)$ EM duality group), the geometry of the CB will be the only tool we have to determine the appropriate subgroup of the flavor outer automorphism group.  In fact we will see that consistency of the geometric deformation will always uniquely determine the action on the flavor symmetry algebra.  We now turn to the simplest such examples.

\subsection{$\Z_k$ gauging of free $U(1)$ theories}\label{U1free}

The generalization of \eqref{N2act} to $C\in\Z_k$ with $k\neq2$ follows by combining the action of a $\s\in\Z_k\subset SL(2,\Z)$ with a similar $\r\in U(1)_R$ and $\vf\in\Out(F)$ as explained in \eqref{N2act1}.  We pointed out already that $\s$ acts non-trivially on $\t$ and is only a symmetry for values of the holomorphic gauge coupling invariant under the $\Z_k$ transformation.  This constraint did not apply to the $\Z_2$ case as $-I$ is in the center of $SL(2,\Z)$ which acts trivially on $\t$.  For $k\neq2$ instead, $\t$ is fixed to a specific $\Z_k$-invariant value given in \eqref{allowedsigma} and thus the daughter theory will always be an isolated SCFT.

For both the $U(1)$ $\cN=4$ and the free $\cN=2$ $U(1)$ gauge theories, the holomorphic gauge coupling is exactly marginal and the above gauging is allowed.  Quotienting the (planar) parent $I_0$ CB geometry by a $\Z_k$ will result in a daughter CB described by a flat $\C/\Z_k$ cone parametrized by $\til u\in \C/\Z_k$ with $\D(\til u)=k$.  From \eqref{allowedsigma} and table \ref{Table:Kodaira}, these are the $IV^*$, $III^*$ and $II^*$ geometries for $k=3,4,6$ respectively.  But as field theories they can be distinguished from other SCFTs with CBs described by the same singularities by their unconventional flavor symmetry groups.  In particular through $\Z_k$ gauging we can ``engineer'' frozen versions of $IV^*$, $III^*$ and $II^*$; see table \ref{Geometries2}.  

\begin{table}[h!]
\small\centering
$\begin{array}{c|c|ccc|c}
\multicolumn{6}{c}{\ccy}\\
\multicolumn{6}{c}{\multirow{-2}{*}{\large Frozen CB geometries of $\Z_k\rtimes U(1)$ gauge theories\ccy}}\\
\hhline{======}
\text{CB geometry}  & \text{field theory content} & \text{flavor} 
& & \text{CB geometry} & \text{flavor} \\
\hhline{===~==}
\multirow{3}{*}{$I_0$}&\multirow{3}{*}{\text{\ free vector multiplet\ }} &\multirow{3}{*}{$\varnothing$}& \xdasharrow{\text{after $\Z_3$ gauging\ \ }} & IV^*  & \varnothing \\
&&& \xdasharrow{\text{after $\Z_4$ gauging\ \ }}  & III^* & \varnothing \\
&&& \xdasharrow{\text{after $\Z_6$ gauging\ \ }} & II^* &\varnothing \\
\hhline{======}
\end{array}$
\caption{\label{Geometries2}
The possible $\Z_k$ discrete gaugings of the $I_0$ singularity.  These provide new frozen singularities which could appear in the deformation pattern of generic deformations.  The three leftmost columns show the CB geometries, field content, and flavor symmetries of the ``parent" free $U(1)$ gauge theories.  The two rightmost columns show the CB geometries and flavor symmetries of the resulting ``daughter" $\Z_k\rtimes U(1)$ gauge theories after discretely gauging the appropriate $\Z_k$ symmetry.
}
\end{table}

Gauging a $\Z_k$ for $k>2$ is not allowed for $U(1)$ gauge theories with charged matter.  The reason is that such a $\Z_k \subset SL(2,\Z)$ is only a symmetry for special values \eqref{allowedsigma} of the gauge coupling, while $U(1)$ gauge theories with matter are IR free.  This means that their couplings vary over the CB, tending to the free value, $\t=i\infty$, at the origin.  Thus IR free theories do not have $\Z_k$ global symmetries in $SL(2,\Z)$ for $k>2$.  But recall that in section \ref{setup} we showed that consistency with $\cN=2$ supersymmetry required the global symmetry to have a non-trivial factor $\s\in SL(2,\Z)$.  

The inconsistency of such gaugings can also be inferred directly from the CB geometry.  To see this, let's go through the same arguments as in the last subsection to attempt to construct a CB geometry for a $\Z_k$ gauging of an $I_n$ singularity.  The $SL(2,\Z)$ monodromy $M$ of this geometry should satisfy $[M^k]=T^n$ and its CB parameter should have scaling dimension $k$.  While there are values of $k$ and $n$ for which such $M$ can be found (e.g., trivially for any $k=n$), all consistent geometries with $\D(\til u)=k$, for $k>2$, have idempotent monodromy matrices (see table \ref{Table:Kodaira}) and thus they cannot satisfy $[M^k]=T^n$.  We thus conclude again that the CB geometry of theories obtained by discrete gauging of a $\Z_k$ for $k>2$ of an IR free theory is not consistent.

%
%

\subsection{Non-lagrangian theories}\label{sec3.3}

We now generalize the construction presented in earlier sections to non-lagrangian theories.  We will show that the extistence of a consistent CB geometry for the daughter theory implies an intricate set of consistency conditions which determines which generators $C=(\r,\s,\vf)$ as in \eqref{N2act1} may be consistently gauged.  

The $C=(\r,\s,\vf)$ generator of the discrete gauge group is inferred from the geometry in the following way.  The possible scale invariant CB geometries fix the subgroup of $SL(2,\Z)$ which acts as a symmetry as shown in \eqref{allowedsigma}.  So for each of these geometries we must select $\s$ to be a non-trivial element of one of these groups.  Say $\s$ generates $\Z_k$.

Next, the corresponding $\r\in U(1)_R$ must then also generate a $\Z_k\subset U(1)_R$, by the argument given in section \ref{setup}.  As explained there, this $\Z_k$ acts on the parent theory CB as a $\Z_r$ with $r=k/\ell$ where $\ell := \D(u)$ is the mass dimension of the parent CB parameter \eqref{rdef}.  This then fixes the CB geometry of the daughter theory in the scale invariant limit (that is, when all mass parameters are set to zero).  In particular, gauging $C$ gives a daughter CB singularity with parameter $\til u$ of scaling dimension $\D(\til u)=r\D(u)$.  As is seen from table \ref{Table:Kodaira}, this uniquely identifies the resulting CB geometry.  

For instance a discrete gauging of the $[IV^*,E_6]$ SCFT can only involve a $\r$ which generates a $\Z_2$ action on its CB, giving a daughter theory CB with a parameter of scaling dimension 6, which is identified as a $II^*$ theory.  Note that this puts a constraint on what $\s$ can be:  since $k=\ell r$, in this case $k=6$, so $\s$ must generate a $\Z_6 \simeq \Z_2\times\Z_3$ which, by \eqref{allowedsigma}, is the whole symmetry subgroup of $SL(2,\Z)$ for this theory.

Next, turn on the most general mass deformation of the parent theory which preserves a $\Z_r$ symmetry of the CB.  This is typically only a subset of the most general allowed mass deformations of the parent theory, and so identifies $\vf\in\Out(F)$ as the outer automorphism of the parent flavor symmetry, $F$, which leaves invariant only those $\Z_r$-preserving masses.  Using the techniques extensively explained in \cite{paper1, paper2}, the deformation pattern of a scale invariant CB singularity under mass deformation allows us to construct a unique consistent Seiberg-Witten geometry.  The SW curve fixes a discrete subgroup (typically the Weyl group) of the flavor symmetry of the resulting theory.  This does not uniquely fix the flavor symmetry algebra, yet it strongly constrains it to a few possibilities as described in \cite{allm1602}.  We can uniquely fix it with the additional requirement that the flavor symmetry algebra of the daughter theory, $F'$, has to be obtained by modding out the known flavor symmetry of the parent theory, $F$, by an action $\G_f$ of its outer automorphism group determined by a choice of $f$ as described below \eqref{OutFact}.  It is a non-trivial result that a solution of the form $F'=F/\G_f$, with $\G_f\subset\Out(F)$, always exists for the flavor symmetry of all daughter theories. This consistency check should be seen as  corroborating evidence for the existence of these theories and it also picks up a particular $\G_f$ and thus a consistent choice for $f$ in \eqref{OutFact}.

Under the action of the $\Z_r$ symmetry on the CB of the parent theory with only the $F'$ mass deformations turned on,  singularities which are located at non-zero values of $u$ which are related by $\Z_r$ phases will be identified upon discrete gauging.   Also, gauging this $\Z_r$ will act on any $I_n$ singularity at the origin of the CB, $u=0$, according to the rules described in sections \ref{U1matt} and \ref{U1free}, and summarized in tables \ref{Geometries} and \ref{Geometries2}.  (Note that the absence of a singularity at $u=0$ corresponds to an $I_0$ ``singularity" in the classification of table \ref{Table:Kodaira}.)

For example, we argued above that there is a single possible $\Z_6$ discrete symmetry of the $[IV^*,E_6]$ SCFT which may be gauged consistently with $\cN=2$ supersymmetry.  Furthermore, we saw that this symmetry acts as the $\Z_2$ generated by $\r: u\mapsto -u$ on the CB of the parent theory, thus leading to a $II^*$ CB geometry upon discrete gauging.  Now, the outer automorphism group of $E_6$ is $\Out(E_6)=\Z_2$,\footnote{Recall that outer automorphisms of simple Lie algebras are just the symmetries of their Dynkin diagrams, so the only non-trivial ones are $\Out(A_{n>1}) = \Out(D_{n>4}) = \Out(E_6) = \Z_2$ and  $\Out(D_4) = S_3$.} so we might expect that the mass deformations of the $E_6$ parent theory which are invariant under $\Out(E_6)$ will preserve a $\Z_2$ symmetry on the CB.  It is not too hard to see that this is, in fact, the case, by using the explicit form \cite{Minahan:1996fg} of the $E_6$ SW curve:  such deformations split the $IV^*$ singularity as $IV^*\to\{{I_1}^8\}$ with four pairs of $I_1$ singularities each located at opposite values of $u$, i.e., at $u=\pm u_i$, $i=1,2,3,4$.  Upon gauging this discrete symmetry, each pair of $I_1$'s is identified with a single $I_1$ in the daughter theory, and the $I_0$ at origin becomes a frozen $I_0^*$, as in the first line of table \ref{Geometries}.  Thus the deformation pattern of the daughter theory is $II^*\to\{I_1^4,I_0^*\}$.  The SW geometry corresponding to precisely such a deformation pattern was constructed in \cite{paper2}, and, furthermore, was found to be invariant under the Weyl group of the $F_4$ exceptional group acting on its mass deformation parameters.   Since a possible action of $\Out(E_6)\simeq\Z_2$ on $E_6$ gives $E_6/\Out(E_6) = F_4$ \cite{Tachikawa:2011ch, Kac:1994IDL}, this is consistent, in a highly non-trivial way, with the above determination of the $C=(\r,\s,\vf)$ generator of the (unique) $\Z_6$ symmetry of the $[IV^*,E_6]$ SCFT which commutes with $\cN=2$ supersymmetry. Notice that the geometry of the daughter theorys could also be interpreted as a $[II^*,D_4\rtimes S_3]$ or $[II^*,U(1)^4\rtimes \G_{F_4}]$ \cite{allm1602}, yet there are no choices of the action of $\Out(E_6)$ which could give either flavor algebras. This shows, as mentioned in passing above, that the analysis of the CB geometry under discrete symmetry not only provides a consistency check for the existence of the daughter theory, but also uniquely identifies the choice of the element of $\Out(F)$ in $\Aut(F)$.

\begin{center}
---\ \ $*$\ \ ---
\end{center}

\begin{figure}[tbp]
\centering
\begin{tikzpicture}
[
auto,
good/.style={rectangle,rounded corners,fill=green!50,inner sep=2pt},
bad/.style={rectangle,rounded corners,fill=red!40,inner sep=2pt},
ugly/.style={rectangle,rounded corners,fill=blue!08,inner sep=2pt},
goodarrow/.style={->,shorten >=1pt,very thick,green!70!black},
badarrow/.style={->,shorten >=1pt,very thick,red},
uglyarrow/.style={->,shorten >=1pt,very thick,blue}
]
\begin{scope}[scale=0.95]
\begin{scope}[yshift=-0cm]
\node at (4.5,-.5) {{\large{\fontfamily{qcs}\selectfont\textsc{$I_0^*$ Series}}}};
\end{scope}
\begin{scope}[yshift=-2cm]
\fill[color=yellow!70, rounded corners] (0,.5) rectangle (9,-.5);
\node at (.6,0) {$II^*:$};
\node (F4) at (2.2,0) [good,align=center] {$F_4$};
\node (D4) at (4.4,0) [ugly,align=center] {$D_4\rtimes S_3$};
\node (4U1) at (7,0) [ugly,align=center] {$U_1^4\rtimes \G_{\!F_4}$};
\end{scope}
\begin{scope}[yshift=-4cm]
\fill[color=yellow!70, rounded corners] (0,.5) rectangle (9,-.5);
\node at (.6,0) {$III^*:$};
\node (B3) at (1.5,0) [good,align=center] {$B_3$};
\node (A3) at (3.5,0) [ugly,align=center] {$A_3\rtimes\mathbb{Z}_2$};
\node (3A1) at (5.5,0) [ugly,align=center] {$A_1^3\rtimes S_3$};
\node (3U1) at (7.5,0) [ugly,align=center] {$U_1^3\rtimes \G_{\!B_3}$};
\end{scope}
\begin{scope}[yshift=-6cm]
\fill[color=yellow!70, rounded corners] (0,.5) rectangle (9,-.5);
\node at (.6,0) {$IV^*:$};
\node (A2) at (3,0) [good,align=center] {$A_2$};
\node (2U1) at (6,0) [ugly,align=center] {$U_1^2\rtimes \Z_2$};
\end{scope}
\begin{scope}[yshift=-8cm]
\fill[color=yellow!70, rounded corners] (0,.5) rectangle (9,-.5);
\node at (.6,0) {$I_0^*:$};
\node (fro) at (4.5,0) [good,align=center] {$\varnothing$};
\end{scope}
\draw[goodarrow] (F4) to [out=225,in=85] (B3);
\draw[badarrow] (F4) to (A3);
\draw[badarrow] (F4) to (3A1);
\draw[badarrow] (F4) to (3U1);
\draw[uglyarrow] (D4) to (B3);
\draw[goodarrow] (D4) to (A3);
\draw[badarrow] (D4) to (3A1);
\draw[badarrow] (D4) to (3U1);
\draw[uglyarrow] (4U1) to (B3);
\draw[uglyarrow] (4U1) to (A3);
\draw[uglyarrow] (4U1) to (3A1);
\draw[goodarrow] (4U1) to [out=315,in=95]  (3U1);
\draw[goodarrow] (B3) to [out=295,in=115] (A2);
\draw[badarrow] (B3) to (2U1);
\draw[goodarrow] (A3) to (A2);
\draw[badarrow] (A3) to (2U1);
\draw[uglyarrow] (3A1) to (A2);
\draw[goodarrow] (3A1) to (2U1);
\draw[uglyarrow] (3U1) to (A2);
\draw[goodarrow] (3U1) to [out=245,in=65]  (2U1);
\draw[goodarrow] (A2) to [out=295,in=105] (fro);
\draw[goodarrow] (2U1) to [out=245,in=75] (fro);
\end{scope}
\begin{scope}[xshift=+9cm,scale=0.95]
\begin{scope}[yshift=-0cm]
\node at (3.5,-.5) {{\large{\fontfamily{qcs}\selectfont\textsc{$I_2^*$ Series}}}};
\end{scope}
\begin{scope}[yshift=-2cm]
\fill[color=yellow!70, rounded corners] (0,.5) rectangle (7,-.5);
\node at (.6,0) {$II^*:$};
\node (G2) at (1.5,0) [good,align=center] {$C_2$};
\node (A2) at (3.5,0) [good,align=center] {$C_1^2\rtimes\Z_2$};
\node (2U1) at (5.5,0) [ugly,align=center] {$U_1^2\rtimes\G_{C_2}$};
\end{scope}
\begin{scope}[yshift=-4cm]
\fill[color=yellow!70, rounded corners] (0,.5) rectangle (7,-.5);
\node at (.6,0) {$III^*:$};
\node (A1) at (2.5,0) [good,align=center] {$C_1$};
\node (U1) at (4.5,0) [good,align=center] {$U_1\rtimes\Z_2$};
\end{scope}
\begin{scope}[yshift=-6cm]
\fill[color=yellow!70, rounded corners] (0,.5) rectangle (7,-.5);
\node at (.6,0) {$I_2^*$};
\node (fro) at (3.5,0) [good,align=center] {$\varnothing$};
\end{scope}
\draw[goodarrow] (G2) to (A1);
\draw[badarrow] (G2) to (U1);
\draw[uglyarrow] (A2) to (A1);
\draw[goodarrow] (A2) to (U1);
\draw[uglyarrow] (2U1) to (A1);
\draw[goodarrow] (2U1) to (U1);
\draw[goodarrow] (A1) to (fro);
\draw[goodarrow] (U1) to (fro);
\draw[goodarrow] (A2) to (fro);
\end{scope}
\end{tikzpicture}
\caption{Green, blue and red arrow label \textit{matching}, \textit{compatible} and \textit{unphysical} RG flows while green and blue backgrounds indicate \textit{good} and \textit{ugly} theories respectively.  While there is always a matching RG flow pattern for all \textit{good} theories in the figure, there are other flows which are necessarily only \textit{compatible} for the \textit{ugly} ones.
\label{I0s}}
\end{figure}
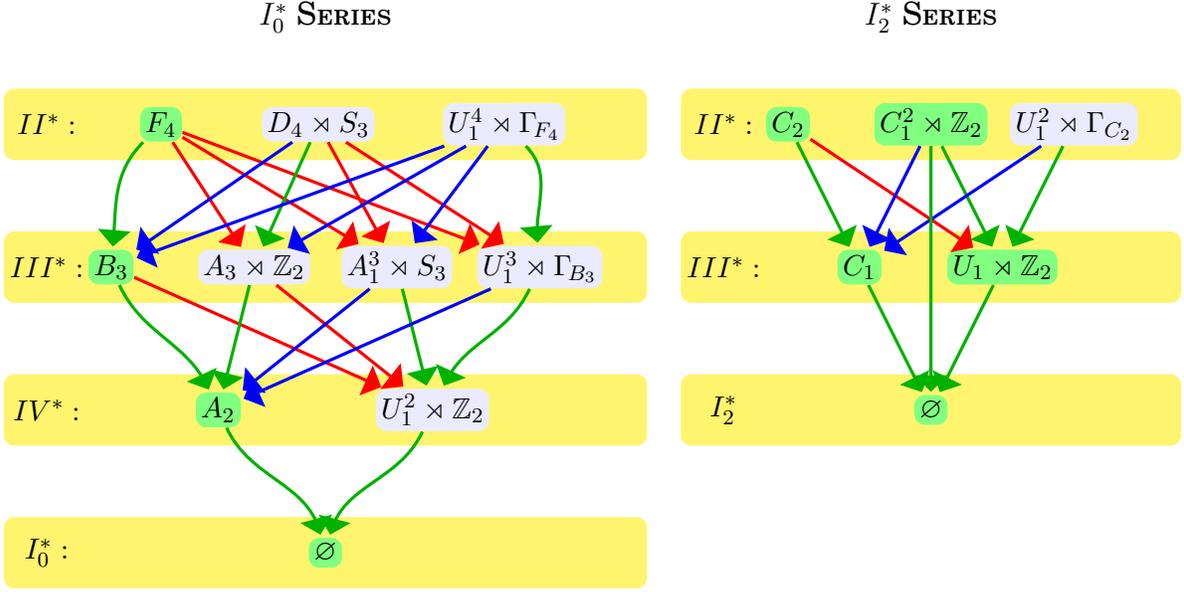

In the rest of this subsection we carry out this kind of argument for every known rank-1 $\cN=2$ SCFT to determine all their possible $\cN=2$ discretely-gauged daughter theories.  The results are summarized in table \ref{tab1}.  Below we organize the discussion into six categories: the $I_0^*$, $I_2^*$, $IV^*$, $III^*$, and $II^*$ series, and $\cN=3$ theories.  The series are named for  the highest-order frozen singularity in their deformation patterns \cite{paper1,paper2,allm1602,Argyres:2016xmc}.  The $\cN=3$ theories are mostly\footnote{Except for one which could be thought of as being the sole member of an ``$I^*_1I_2$ series".} special cases of the other series, but because of their enhanced supersymmetry require a separate discussion.  Theories in the same series are connected by RG flows, shown as vertical arrows in table \ref{tab1}.  They also have to satisfy extra checks arising from the requirement of consistency of flavor symmetry-breaking under RG flows: the breaking of the flavor symmetry algebra along RG flow directions should match the flavor symmetry algebra assignment which can be read off from the singularity structure along that RG direction.  Following the terminology introduced in \cite{allm1602}, RG flows can be {\it matching}, {\it compatible} or {\it unphysical}.  The results of this RG flow analysis for the $I_0^*$ and $I_2^*$ series are reported in figure \ref{I0s}; those for the $IV^*$ and $III^*$ series were already reported in \cite{allm1602}; and those for the remaining series are trivial.  For more details and a systematic explanation of the RG flow consistency condition we refer the reader to \cite{paper2, allm1602}.



\paragraph{\emph{I}$^*_{\bf 0}$ series.} 


These are the daughter theories which flow to a frozen $I_0^*$ CB singularity upon generic relevant deformation.  Aside from the discrete gauging construction outlined in the beginning of this section, there is no lagrangian interpretation of a frozen $I_0^*$ singularity, suggesting that the only consistent interpretation of theories in this series is via discrete gauging.\footnote{The existence of a rank-0 interacting SCFT with appropriate central charge values and a flavor symmetry containing an $A_1$ subalgebra with an empty commutant, would invalidate that statement since we could gauge such an $A_1$ factor to build a non-lagrangian version of a frozen $I_0^*$ theory.  A more detailed discussion of this possibility can be found in \cite{paper2}; we will not consider this possibility any further here.}  The frozen $I_0^*$ can then be interpreted as a $\Z_2$ gauging of a free vector multiplet with $I_0$ CB geometry, so for all the theories in this series, the action of the discretely gauged group on the CB is a $\Z_2$.  These are therefore those theories in the $\Z_2$ column of table \ref{tab1} with arrows leading to the free $[I_0^*,\varnothing]$ $\cN=2$ $O(2)$ gauge theory --- i.e., the $[II^*,F_4]$, $[III^*,B_3]$, and $[IV^*,A_2]$ theories.  

(Also, the bottom two rows of the $\Z_2$ and $\green{\til\Z_2}$ columns of table \ref{tab1} show free  theories which flow to $[I_0^*,\varnothing]$.  They are the $\cN=4$ $O(2)$ gauge theory $[I_0^*,C_1\chi_0]$, discussed previously in \cite{gr1512}, and the theory of an $\cN=2$ $O(2)$ gauge theory with a decoupled hypermultiplet, denoted by $[I_0^*,\chi_0]\times\H$.  We will discuss these theories in section \ref{N3}.)

The gauged subgroup of $SL(2,\Z)$, as explained above, is a $\Z_{2 \D(u)}$, where $\D(u)$ is the scaling dimension of the CB parameter of the parent theory.  From table \ref{tab1} we can then read off the $SL(2,\Z)$ actions as $\Z_6$, $\Z_4$ and $\Z_3$ for the $[II^*,F_4]$, $[III^*,B_3]$ and $[IV^*,A_2]$ cases, respectively.  The discrete gauging of the $[IV^*,E_6]$ and $[I_0^*,D_4]$ parent theories, enforces also the gauging of a $\Z_2$ outer automorphism of the flavor symmetry algebra, giving daughter theories with $F_4\cong E_6/\Z_2$ and $B_3\cong D_4/\Z_2$ flavor symmetries.  Perhaps unexpectedly, the discrete gauging of the $[IV,A_2\chi_{1/2}]$ does not act on the flavor symmetry algebra, but only on the $\chi_{1/2}$ chiral deformation of the $IV$ singularity, freezing it.  It is in fact remarkable that the generic deformation of the $IV$ singularity with $\chi_{1/2}=0$ fully splits $IV\to\{{I_1}^4\}$, but nevertheless preserves a $\Z_2$ CB symmetry locating the four $I_1$'s at pairwise opposite points, $u=\pm u_j$, $j=1,2$.

In figure \ref{I0s} we report the RG-flow analysis for the $I^*_0$ series. The only {\it matching} interpretation of the series, in the language introduced in \cite{allm1602}, is the one consistent with the outer automorphism action explained above. 

\paragraph{\emph{I}$^*_{\bf 2}$ series.} 

This series only contains three theories, also appearing in the $\Z_2$ and $\green{\til\Z_2}$ columns of table \ref{tab1}.  In the $\Z_2$ column, one is a $[II^*,C_2]$ theory, daughter of the $[IV^*,C_2U_1]$ theory, and the other is a $[III^*,C_1]$ theory, daughter of the $[I_0^*,C_1\chi_0]$ theory.
The frozen $I_2^*$ in the generic deformation of the daughter theories comes from a $\Z_2$ gauging of an $I_4$ located at the origin of the parent theory, while pairs of the remaining $I_1$'s in the parent theory are identified.  The $[IV^*, C_2U_1] \to [II^*, C_2]$ is a $\Z_6$ discrete gauging which acts as a $\Z_2$ on the parent CB, and the $[I_0^*, C_1\chi_0] \to [III^*, C_1]$ is a $\Z_4$ discrete gauging which acts as a $\Z_2$ on the parent CB.  In the first case the discretely-gauge group includes an action of the outer automorphism group on the flavor symmetry: $C_2U_1/\Out(C_2 U_1) = C_2 \times (U_1/\Out(U_1)) = C_2$. In the second case the $SL(2,\Z)$ action freezes the marginal $\Chi_0$ deformation (i.e., fixes the gauge coupling to $\t=i$) and does not act on the flavor group;  indeed, $\Out(C_1)$ is trivial, so $C_1/\Out(C_1)=C_1$.  Finally, the RG-flow analysis for the $I_2^*$ series shown in figure \ref{I0s} shows that these theories correspond to {\it matching} flows.

However, there is an ambiguity in determining the $\Z_4$ symmetry of the $[I_0^*,C_1\chi_0]$ which can be gauged.  The reason is that the $[I_0^*,C_1\chi_0]$ theory is the (lagrangian) $SU(2)$ $\cN=4$ SYM theory, which has an enhanced supersymmetry, and so has more than one discrete symmetry group that preserves an $\cN=2$ supersymmetry.  In fact, we will argue in the next subsection (on $\cN=3$ theories) that there is a consistent gauging of a second $\Z_4$ which gives a $[III^*,U_1\rtimes\Z_2]$ daughter theory with $\cN=3$ supersymmetry.  Note that, according to figure \ref{I0s} a flow from the $[II^*,C_2]$ theory to a $[III^*,U_1\rtimes\Z_2]$ theory is {\it unphysical}, so the latter theory must belong to a separate RG flow.  This is the $I^*_2$-series theory shown in the $\green{\til\Z_2}$ column in table \ref{tab1}.

\paragraph{\emph{IV}$^*$ series.} 

Theories in this series are those daughter theories that flow to frozen $IV^*$ singularities under generic deformation.  Since a $[IV^*,\varnothing]$ singularity only appears as the the result of a $\Z_3$ gauging of an $I_0$ (free vector multiplet) theory, the theories in this series appear in the $\Z_3$ and $\green{\til\Z_3}$ columns of table \ref{tab1}:  a $\Z_3$ action on the deformed CB of their parent theories transforms the $I_0$ at the origin into the frozen $IV^*$ at the end of the RG flow.  (The $Q=1$ and $Q=\sqrt2$ subscripts on the frozen $IV^*$ theories are to distinguish the unit of normalization of electric and magnetic charges in the low energy theory on the CB; see \cite{paper1} for a discussion.)  

The $[II^*,G_2]$ theory is obtained from the $[I_0^*,D_4\chi_0]$ theory (i.e., $N_f=4$ $SU(2)$ sQCD) by picking a $\Z_3 \subset S_3 \cong \Out(D_4)$ flavor symmetry action.   The $[III^*,A_1]$ is obtained from the $[III,A_1\chi_{2/3}]$ by freezing the $\chi_{2/3}$ chiral deformation; no flavor symmetry action is required. This is compatible with the remarkable fact that a generic deformation of the $III$ with $\chi_{2/3}=0$ splits it into three $I_1$ singularities which are always at the vertices of an equilateral triangle, thus preserving the $\Z_3$ symmetry of the CB geometry. 
Furthermore, these flavor assignments give {\it matching} RG flow flows, according to the RG-flow analysis for this series presented in \cite{allm1602}.


Note that the frozen $IV^*$ series was already considered and analyzed in \cite{allm1602} but with different conclusions for the correct flavor assignments for the $II^*$ and $III^*$ theories, namely $[II^*,A_2\rtimes \Z_2]$ and $[III^*,U_1\rtimes\Z_2]$, and these in fact appear in the ``parent" column of table \ref{tab1}.  This is not a contradiction:  the theories analyzed in \cite{allm1602} did not come from discrete gauging and they are thus different from the $[II^*,G_2]$ and $[III^*,A_1]$.  The fact that a single CB geometry, associated with a given deformation pattern, can correspond to multiple theories is due to the fact that the frozen $IV^*$ allows for both a discretely gauged and a non-discretely gauged interpretation.  This is not surprising since we know already of the example of the frozen $I_1^*$ geometry which exists both as $\Z_2$ discretely gauged version of a $U(1)$ theory with a single hypermultiplet with charge $\sqrt{2}$ and as an $SU(2)$ gauge theory with a single half-hypermultiplet in the spin $3/2$ representation \cite{paper1}.  

\paragraph{\emph{III}$^*$ series.} 

Theories in this series are those daughter theories that flow to frozen $III^*$ singularities under generic deformation.  Since a $[III^*,\varnothing]$ singularity only appears as the the result of a $\Z_4$ gauging of an $I_0$ (free vector multiplet) theory, the theories in this series appear in the $\Z_4$ and $\green{\til\Z_4}$ columns of table \ref{tab1}:  a $\Z_4$ action on the deformed CB of their parent theories transforms the $I_0$ at the origin into the frozen $III^*$ at the end of the RG flow.    

The $[II^*,B_1]$ theory is obtained from the $[IV,A_2\chi_{1/2}]$ theory by freezing the $\chi_{1/2}$ chiral deformation, and by dividing by the parent flavor symmetry by $\Z_2\cong \Out(A_2)$.    
Furthermore, this flavor assignment gives a {\it matching} RG flow, according to the RG-flow analysis for this series presented in \cite{allm1602}, although in this case this is a trivial check.


\paragraph{\emph{II}$^*$ series.}

The remaining theories are either green, blue or red in table \ref{tab1}.  The green and blue theories correspond to theories with enhanced supersymmetry and will be discussed in the next subsection.  The theories in red are instead somewhat more speculative than the others because they are frozen and so cannot be connected to any other $\cN=2$ theory through an $\cN=2$ RG flow.  Thus there are very few checks available to give convincing evidence that they actually exist as physical theories.  The CBs of these theories are both $[II^*,\varnothing]$ singularities, one obtained by a $\Z_5$ CB action on the $[II,\chi_{4/5}]$ theory, and the other by a $\Z_2$ CB action of the $\cN=3$ supersymmetric $[IV^*,U_1]$ theory.  In the former case the $\Z_5$ $U(1)_R$ action freezes the $\chi_{4/5}$ chiral deformation of the $II$ while its empty flavor symmetry ``carries over" to form a frozen $II^*$.  In the latter case the $U(1)$ flavor symmetry is frozen by the discrete gauging procedure.  In fact, as there is no non-trivial mass deformation of the $[IV^*,U_1]$ invariant under the $\Out(U_1)\cong\Z_2$, this is compatible with the fact that any non-zero value of the mass associated to the $U(1)$ flavor of the $IV^*$ splits $IV^*\to\{ I_1,I_1^*\}$ so the only $\Z_2$ symmetric mass deformation is the trivial one. 

\subsection{$\cN=3$ theories}\label{N3}

Discretely gauging $\cN=4$ supersymmetric parent theories --- the blue theories in the ``parent" column of table \ref{tab1} --- deserves a separate discussion.  In this case the R-symmetry action of the discrete group can be embedded in the $\cN=4$ $SO(6)_R$ R-symmetry group, as was briefly reviewed at the beginning of section \ref{setup}.  When combined with the appropriate $SL(2,\Z)$ action, there can be more than one $\Z_k$ symmetry group for a given $k$ preserving $\cN=2$ supersymmetry, and therefore more than one daughter theory with the same CB geometry.  In particular, we will find examples of this for $k=3, 4$ and $6$, and in each case there will be one $\Z_k$ which preserves precisely $\cN=2$ supersymmetry, and another which preserves $\cN=3$ supersymmetry.  The $\cN=2$ actions are shown in the $\Z_k$ columns of table \ref{tab1}, while the $\cN=3$ actions are indicated in the  $\green{\til \Z_k}$ columns and will be introduced below.

There are two rank-1 $\cN=4$ CB geometries, the $I_0$ one corresponding to a free $\cN=4$ vector multiplet, and the $I^*_0$ one corresponding to an $\cN=4$ $SU(2)$ SYM theory. The free $\cN=4$ vector multiplet is, from an $\cN=2$ perspective, a free $\cN=2$ vector multiplet plus a massless neutral hypermultiplet.  As said, its CB is described by an $I_0$ ``singularity" with an arbitrary value of $\t$, and an $SU(2) \simeq C_1$ flavor symmetry acting on the hypermultiplet.   It thus appears as the $[I_0,C_1\chi_0]$ theory in the first column of table \ref{tab1}.  Its $\cN=4$ $[I_0^*,C_1\chi_0]$ daughter from a $\Z_2$ gauging is the $O(2)$ $\cN=4$ theory constructed in \eqref{N2act4} and discussed at length in \cite{gr1512}, while the three $\cN=3$ theories obtained as discrete gaugings of the $[I_0,C_1\chi_0]$ theory were previously constructed in \cite{gr1512, Aharony:2016kai}. 

Similarly, the $\cN=4$ $SU(2)$ SYM theory has, from an $\cN=2$ perspective, a CB described by an $I_0^*$ singularity with a marginal $\chi_0$ coupling, and a $C_1$ flavor symmetry, so appears as the $[I_0^*,C_1\chi_0]$ theory in the first column of table \ref{tab1}.  In fact, it appears twice, once in the series ending in an $[I_4,\varnothing]$ and once in the series ending in an $[I_2,\varnothing]$.  These refer to two different CB geometries under deformation.  The $I_4$ series theory has deformation pattern $[I_0^*,C_1\chi_0] \to \{ I_4, I_1, I_1\}$ while the $I_2$ series theory has deformation pattern $[I_0^*,C_1\chi_0] \to \{ I_2, I_2, I_2\}$.  These two CB geometries are related by a 2-isogeny of their elliptic fibers and so seem to differ from one another only by a choice of normalization of the quantization unit of their electric and magnetic charges under the low energy $U(1)$ gauge group \cite{paper2}.  However, surprisingly, we will see that they have different behaviors under discrete gauging.   

\subsubsection{$\cN=2$ and $\cN=3$ preserving discrete gauging ($\Z_r$ vs. $\green{\til \Z_r}$)}

In order to introduce the already mentioned $\cN=3$ preserving $\green{\til \Z_r}$ action, it is instructive to start with an example and re-examine the $\Z_4$ gauging of the $I_4$-series $[I^*_0,C_1\Chi_0]$ theory.  (This is the $[III^*,C_1]$ daughter theory we described in the last subsection in the $I_2^*$-series paragraph.)  We will argue that this is only one of two consistent $\Z_4$ gaugings of this theory, in fact a $\cN=3$ preserving gauging is also allowed.  Let's first run through the discrete gauging argument in order to clearly identify where the ambiguity arises.  

The $\Z_2$ action on the CB of the parent $[I_0^*,C_1\chi_0]$ theory should be accompanied by the action of a $\Z_4\subset SL(2,\Z)$ symmetry in order to preserve $\cN=2$ supersymmetry, since $4=2\cdot\D(u)$.  This fixes the gauge coupling, which is the marginal chiral $\chi_0$ deformation of the parent theory, to $\t=i$.  Then the explicit form of the $[I_0^*,C_1\chi_0]$ theory's SW curve at $\t=i$ (given in \cite{paper2}) shows that its CB is invariant under a $\Z_2$ action for arbitrary $C_1$ mass deformations; see appendix \ref{A2}.  Thus the daughter theory should have a $III^*$ CB geometry with a rank-1 flavor symmetry with a $\Z_2$ Weyl group (i.e., a dimension-2 mass invariant).  Indeed, such a deformed CB branch geometry was found in \cite{paper1, paper2}, and, as explained in \cite{allm1602}, it can consistently have either a $[III^*,C_1]$ or a $[III^*,U_1\rtimes\Z_2]$ flavor symmetry.  But since there is no action of the $\Z_2$ on the parent theory's mass parameter, its $C_1$ flavor group should not be divded by any outer automorphism, so the daughter theory should be the $[III^*,C_1]$ theory.

This analysis, however, made the assumption that the parent theory has only $\cN=2$ supersymmetry, while, in fact, it has $\cN=4$ supersymmetry.  This permits more latitude in constructing discrete symmetry groups which preserve (at least) $\cN=2$ supersymmetry.
The amount of supersymmetry preserved by various discrete symmetries of this theory can be analyzed following \cite{gr1512}.   If $\s$ generates the $\Z_4\subset SL(2,\Z)$ symmetry subgroup of the S-duality group\footnote{More properly, the S-duality group of the $\cN=4$ $SU(2)$ SYM is an index 3 subgroup $\G^0(2) \subset PSL(2,\Z)$ of which $\s$ generates a $\Z_2$ subgroup; the correct discussion will be given below.} of the $\cN=4$ theory at $\t=i$, the chiral $\cN=4$ supercharges transform as $\s: Q^i_\a \mapsto e^{-i\pi/4} Q^i_\a$, where $i\in{\bf 4}$ of $SO(6)_R$, as in \eqref{SLtr1}.  So, in order to preserve at least $\cN=2$ supersymmetry, we need to pick a generator $\r\in SO(6)_R$ so that under the combined action of $(\r,\s)$ at least two of the supercharges are left invariant.  Up to conjugation by elements of $SO(6)_R$, $\r$ can be chosen to be in the maximal torus of $SO(6)_R$, so can be represented by a simultaneous rotation,
\begin{align}\label{N4rho}
\r \simeq (e^{i\psi_1},e^{i\psi_2},e^{i\psi_3}) ,
\end{align}
in three orthogonal planes in $\R^6\simeq \C^3$.  The four chiral supercharges transform under this rotation by the phases $\{ e^{i(\psi_1+\psi_2+\psi_3)/2}, e^{i(\psi_1-\psi_2-\psi_3)/2}, e^{i(-\psi_1+\psi_2-\psi_3)/2}, e^{i(-\psi_1-\psi_2+\psi_3)/2} \}$.  For $\r$ to generate a $\Z_4$, the $\psi_a$ must all be multiples of $\pi/4$.  Then, up to the action of the Weyl group of $SO(6)_R$ (which permutes the $\psi_a$ and shifts any pair of them by $\pi$), there are just two inequivalent solutions for a $\r$ as in \eqref{N4rho} which preserve at least two supersymmetries:
\begin{align}\label{N4rhosols}
\Z_4:\qquad (a) \quad
\r_a := 
\left( i, 1, 1\right)
\qquad \text{and}\qquad 
(b) \quad
\r_b := 
\left( i, i, -i\right).
\end{align}
Combined with the $\s$ action on the supercharges, it follows that solution $(a)$ preserves $\cN=2$ supersymmetry (by leaving $Q^i_\a$ for $i=1,2$ invariant), while solution $(b)$ preserves $\cN=3$ supersymmetry (by leaving $Q^i_\a$ for $i=1,2,3$ invariant). 

From their action on the supercharges it follows that a $\r\simeq(e^{i\psi},1,1)$ rotation is in the $U(1)_R$ factor of the R symmetry of the $\cN=2$ algebra preserved by solution $(a)$, while a $\r\simeq(1,e^{i\psi},e^{i\psi})$ is in the maximal torus of the $SU(2)_R$ factor of its R symmetry.   So $\r_a\in U(1)_R$ and the commutant of $U(1)_R\times SU(2)_R$ in $SO(6)_R$ is an $SU(2)_F$.  Thus gauging $(\r_a,\s)$ preserves a $C_1\simeq SU(2)_F \subset SO(6)_R$ flavor symmetry, as expected from our earlier arguments.  

However, the same analysis applied to the $\r_b$ solution shows that $\r_b \not\in U(1)_R$ for any choice of $\cN=2$ subalgebra of the $\cN=3$ supersymmetry which it preserves.  With respect to  any $\cN=2$ subalgebra, the parent $\cN=4$ R symmetry decomposes as $SO(6)_R \supset U(1)_R\times SU(2)_R\times SU(2)_F$, as in the previous paragraph.  But \eqref{N4rhosols} implies $\r_b\in U(1)_R \times SU(2)_F$ in such a way that the commutant of $\r_b$ and $U(1)_R\times SU(2)_R$ in $SO(6)_R$ is just a $U(1)_F\subset SU(2)_F$.  Thus, by gauging $(\r_b,\s)$, the $\cN=2$ flavor algebra is reduced to $U(1)$, contrary to our earlier arguments.  There is no contradiction with those arguments, however, since in this case $\r_b$ does not generate a subgroup of the $\cN=2$ $U(1)_R$ symmetry.\footnote{Note that $\r_b$ \emph{does} generate a subgroup of the $\cN=3$ $U(1)_R$ symmetry, since elements of the form $\r \simeq (e^{i\psi}, e^{i\psi}, e^{-i\psi})$ rotate $Q^i_\a$ with $i=1,2,3$ by a common phase.} The gauging of $(\r_b,\s)$ is what we call $\green{\til \Z_4}$ above.

The actions of $\r_a$ and $\r_b$ on the $\cN=4$ moduli space can also be easily worked out.  Denote the six real adjoint scalars in the $\cN=4$ vector multiplet by $\vf^I_A$ where $I\in{\bf 6}$ of $SO(6)_R$ and $A\in{\bf 3}$ of the $SU(2)$ gauge group.  These can be combined into three complex adjoint scalars $\f^a_A := \vf^{2a-1}_A + i \vf^{2a}_A$ for $a=1,2,3$.
Then $\r$ in \eqref{N4rho} acts as $\r:\f^a_A \mapsto e^{i\psi_a} \f^a_A$ on the adjoint scalars.  This implies that with respect to the $\cN=2$ algebra fixed by $\r_a$, $\f^1_A$ is the complex adjoint scalar in the $\cN=2$ $SU(2)$ vector multiplet while $(\f^2_A,\bar{\f^3_A})$ are the scalars in the adjoint hypermultiplet.  The same therefore is also true for solution $(b)$ with respect to the choice of $\cN=2$ subalgebra generated by $Q^1_\a$ and $Q^2_\a$.  The moduli space of the parent $SU(2)$ $\cN=4$ SYM theory is parameterized by the vevs of the holomorphic gauge invariant ``meson" fields $M^{(ab)} := \sum_A \f^a_A \f^b_A$, subject to the relations $M^{ab} M^{cd} = M^{ac} M^{bd}$ following from the usual F- and D-term equations.  This is equivalent to a $\C^3/\Z_2$ orbifold; the $\Z_2$ is the residual identification by the Weyl group of the $SU(2)$ gauge group.  The $M^{11}$ vev then parametrizes the CB with respect to the $Q^1_\a$, $Q^2_\a$ $\cN=2$ subalgebra, $M^{12}$ and $M^{13}$ parameterize the mixed branch directions, and the rest are coordinates on the Higgs branch.  $\r_a$ and $\r_b$ both act by $M^{11}\mapsto -M^{11}$ on the CB, giving the same $III^*$ singularity, but have different actions on the Higgs and mixed branches.

\begin{center}
---\ \ $*$\ \ ---
\end{center}

The above computation of the $\Z_k$ subgroups of the $\cN=4$ $SO(6)_R$ R-symmetry which preserve $\cN=2$ supersymmetry (when combined with a $\Z_k\subset SL(2,\Z)$ action) generalizes immediately to all $k$.  The same argument as in the paragraph containing \eqref{N4rho} and \eqref{N4rhosols} leads to two solutions for all $k$:
\begin{align}\label{N4rhosols-k}
\Z_k:\   
\begin{cases}
\r_a &:=\  
\left( e^{2\pi i/k}, 1, 1\right)\\
\r_b &:=\  
\left( e^{2\pi i/k}, e^{2\pi i/k}, e^{-2\pi i/k}\right)
\end{cases}
, \qquad k\in\{2,3,4,6\}.
\end{align}
Combined with the $\s\in\Z_k\subset SL(2,\Z)$ action on the supercharges \eqref{SLtr1}, it follows that $\r_a$ preserves only an $\cN=2$ supersymmetry and $\r_b$ preserves an $\cN=3$ supersymmetry except for $k=2$, where it preserves the whole original $\cN=4$ supersymmetry. 

The $\r_b$ solution is, in fact, the R-symmetry action described by Garc\'ia-Etxebarria and Regalado in \cite{gr1512} and is what we call $\green{\til \Z_k}$ in table \ref{tab1}. When applied to the $[I_0,C_1\chi_0]$ parent theory in the ``parent" column of table \ref{tab1} (i.e., a free $\cN=4$ vector multiplet), the resulting blue and green daughter theories in the table are the $O(2)$ $\cN=4$ theory and some of the $\cN=3$ theories constructed in \cite{gr1512} (they are the $k=2,3,4,6$ with $\ell=1$ theories in the notation of \cite{Aharony:2016kai}). 

If we discretely gauge in this same parent theory the $\Z_k$ with $\r_a$ generator, instead, we find a series of $\cN=2$ daughter theories, denoted in table \ref{tab1} as $[K]\times\H$ for $K\in\{I_0^*,IV^*,III^*,II^*\}$.  This is easy to understand:  the parent $[I_0,C_1\chi_0]$ theory is, as an $\cN=2$ theory, just a free vector multiplet (giving an $[I_0]$ singularity) plus a free neutral massless hypermultiplet (denoted by $\H$).  The $\r_a$-gauging acts on the vector multiplet in the way described earlier in section \ref{U1free} to give the frozen $[K]$-type CB geometries, and does nothing to the hypermultiplet, leaving its Higgs fiber, $\H$, unaffected.

\subsubsection{New $\cN=3$ theories}

Let's now apply the $\Z_k$, with CB action given by $\r_a$, and $\green{\til \Z_k}$, with a CB action given by $\r_b$, discrete gaugings, or , to the $[I^*_0,C_1\chi_0]$ parent theory in the ``parent" column of table \ref{tab1} (i.e., an $\cN=4$ $SU(2)$ super YM theory).  In this case we find some surprises.  Since $\D(u)=2$ for the $I_0^*$ CB parameter, the $\Z_k$ discrete symmetry acts only as $\Z_r$ with $r=k/2$ on the CB.  Thus there are only two possibilities: $k=4$  or $k=6$.  In the $k=4$ case the $\chi_0$ marginal deformation is frozen at $\t=i$, while for the $k=6$ case it is frozen at $\t=e^{i\pi/3}$.  Our analysis then predicts that in the $\r_a$ case the daughter theories will be $\cN=2$ SCFTs with CB geometries $[III^*,C_1]$ (for $k=4$) and $[II^*,C_1]$ (for $k=6$).  Likewise, in the $\r_b$ case the daughter theories will be $\cN=3$ SCFTs with CB geometries $[III^*,U_1\rtimes\Z_2]$ (for $k=4$) and $[II^*,U_1\rtimes\Z_2]$ (for $k=6$).  (See \cite{allm1602} for an explanation of the $\Z_2$ factors in their flavor symmetries.)  

The first surprise is that these $\cN=3$ theories do not appear on the list of $\cN=3$ theories found in \cite{gr1512, Aharony:2016kai} by a string S-folding construction.  In fact, they are the unshaded $k=4,6$ with $\ell=2$ entries in table (2.13) of \cite{Aharony:2016kai}.  Since the parent theory is a lagrangian theory, our explicit identification of $\Z_4$ and $\Z_6$ global symmetries which commute with three supercharges would seem to guarantee the existence of these $\cN=3$ theories upon gauging these symmetries.  However, the global symmetries in question include the action of symmetry subgroups of the group of S-duality transformations, and these only occur at strong coupling.  So one might worry that there is some subtlety having to do with the existence of these symmetries that cannot be seen at weak coupling.  Indeed, just such a subtlety is the second surprise, which we turn to now.

As we already mentioned, there are two distinct CB geometries describing consistent deformations of the $I_0^*$ singularity with one mass parameter.  One is the $[I^*_0,C_1\chi_0]$ entry in the $I_4$ series RG flow (i.e., the twelfth entry from the top in the ``parent" column of table \ref{tab1}), and the other is the $[I^*_0,C_1\chi_0]$ entry in the $I_2$ series RG flow (i.e., the fifth entry from the bottom in the ``parent" column of table \ref{tab1}).  The $I_4$-series version splits as $[I^*_0,C_1\chi_0] \to \{ I_4, I_1, I_1\}$ upon turning on the mass deformation, while the $I_2$-series version splits as $[I^*_0,C_1\chi_0] \to \{ I_2, I_2, I_2\}$.

Let's first discuss the $[I^*_0,C_1\chi_0] \to \{ I_4, I_1, I_1\}$ case. As we noted in \cite{paper1} --- see especially the last paragraph of section 5.3 --- the $I_4$-series curve describes the $SU(2)$ $\cN=4$ theory with S-duality group $\G^0(2) \subset PSL(2,\Z)$, i.e., the index-3 subgroup generated by $T^2$ and $STS$.\footnote{The S-duality group of the $\SU(2)$ SYM theory is a subgroup of $PSL(2,\Z)$, not $SL(2,\Z)$, since the center of $SL(2,\Z)$ is part of the gauge group, e.g., on the moduli space its action on dyon charges is just that of the Weyl group.}
The fundamental domain of $\G^0(2)$ has two weak-coupling cusps and a $\Z_2$ orbifold point.  We can pick the fundamental domain so that one cusp is at $\t=0$ (with $2\pi$ theta angle identification), the other is at $\t=i\infty$ (with $4\pi$ theta angle identification), and the $\Z_2$ orbifold point is at $\t = i\pm1$ (which are identified by $T^2$).  The $\t=0$ limit is the $SU(2)$ theory and the $\t=i\infty$ limit is the GNO-dual $SO(3)$ theory.  
The $\t=i+1$ orbifold point is fixed by $\s := T^2 S T S = \left(\begin{smallmatrix}1&-2\\1&-1\end{smallmatrix}\right)$ which satisfies $\s^2=I$ (in $PSL(2,\Z)$, though not in $SL(2,\Z)$) and which is an element of the S-duality group.  It thus generates a $\Z_2$ global symmetry of the theory, which acts, according to \eqref{SLtr}, as $Q_\a^i \mapsto e^{-i\pi/4} Q_\a^i$.  Note the difference from the action \eqref{SLtr1} which applied to the case where the EM-duality group was $SL(2,\Z)$.  In general, when the S-duality group is (a subgroup of) $PSL(2,\Z)$, the action on the supercharges becomes
\begin{align}\label{PSLtr1}
PSL(2,\Z) \supset\Z_r \ni \s:\  Q_\a^i \to {\rm e}^{-i\pi/(2r)}Q_\a^i.
\end{align}
Then our previous arguments for the discrete symmetry which preserves at least two supersymmetries go through with $\r_a$ and $\r_b$ as in \eqref{N4rhosols-k} with $k=2r$.

Thus, we have identified two $\Z_4$ global symmetries of the $I_4$-series $[I_0^*,C_1\chi_0]$ theory at the value $\t=1+i$ of its marginal coupling,
\begin{align}\label{}
C_a := (\r_a,\s) 
\quad\text{and}\quad
C_b := (\r_b,\s)
\quad \in \quad
SO(6)_R \times PSL(2,\Z),
\end{align}
with $C_a$, $C_b$ preserving only an $\cN=2$, $3$ supersymmetry, respectively.  Gauging these two symmetries then gives the $[III^*,C_1]$ and $[III^*,U_1\rtimes\Z_2]$ theories, respectively, as described above.  

Note that neither of $\pm ST$ (or any of their conjugates) are elements of $\G^0(2)\subset PSL(2,\Z)$, so they do not generate a symmetry of the theory at $\t=e^{2\pi i/3}$ (which they fix), and so there is not an identification of the theory at $\t=e^{2\pi i/3} +\e$ with the theory at $\t=e^{2\pi i/3} + e^{2\pi i/3}\e + \cO(\e^2)$.  Indeed there is no $\Z_3$ orbifold point of the $\G^0(2)$ fundamental domain.  Since it has no $\Z_3$ S-duality symmetry, there is no $\Z_6$ global symmetry of the $I_4$-series $[I_0^*,C_1\chi_0]$ theory, and so no possible daughter $II^*$ theories with $\cN=2$ and $\cN=3$ supersymmetry.

Now let's turn to a discussion of the $I_2$-series curve which describes a subtly different version of this theory.  In the weak-coupling limit it appears to be identical to an $\cN=4$ $\SU(2)$ SYM theory:  their SW geometries are related by a 2-isogeny of their elliptic fibers, constructed explicitly in \cite{paper2}, which does not affect the low-energy observables or the BPS spectrum.  This 2-isogeny identification is reflected in a change in the charge quantization unit by a factor of $\sqrt2$ together with a rescaling of the marginal coupling $\t$ by a factor of $2$.  Although this factor of two is just a change of variables in the weak coupling limit, it cannot be removed by a change of variables for all values of $\t$ without changing the global properties of the S-duality identifications of the low energy theory qualitatively.  In particular, the $I_2$ series SW curve (first in found in \cite{sw2} and reviewed in \cite{paper2}) is invariant under the full $PSL(2,\Z)$ S-duality group, and not just a subgroup as in the $I_4$-series case.  

This difference has concrete consequences for the allowed discrete gaugings which preserve an $\cN=2$ supersymmetry.  In particular, the S-duality group, $PSL(2,\Z)$, of the $I_2$-series theory contains both a $\Z_2$ subgroup (generated by $S$) and a $\Z_3$ subgroup (generated by $ST$), it has both $\Z_4$ and $\Z_6$ symmetries which commute with enough supersymmetries.  (Equivalently, the fundamental domain of $PSL(2,\Z)$, unlike that of $\G^0(2)$, has both a $\Z_2$ and a $\Z_3$ orbifold point.)  
This means then that these can be combined with $\Z_4$ and $\Z_6$ subgroups of $SO(6)_R$ generated by $\r_a$ or $\r_b$ given in \eqref{N4rhosols-k} to construct both $\cN=2$ daughter $[III^*,C_1]$ and $[II^*,C_1]$ theories, as well as $\cN=3$ daughter $[III^*,U_1\rtimes\Z_2]$ and $[II^*,U_1\rtimes\Z_2]$ theories.

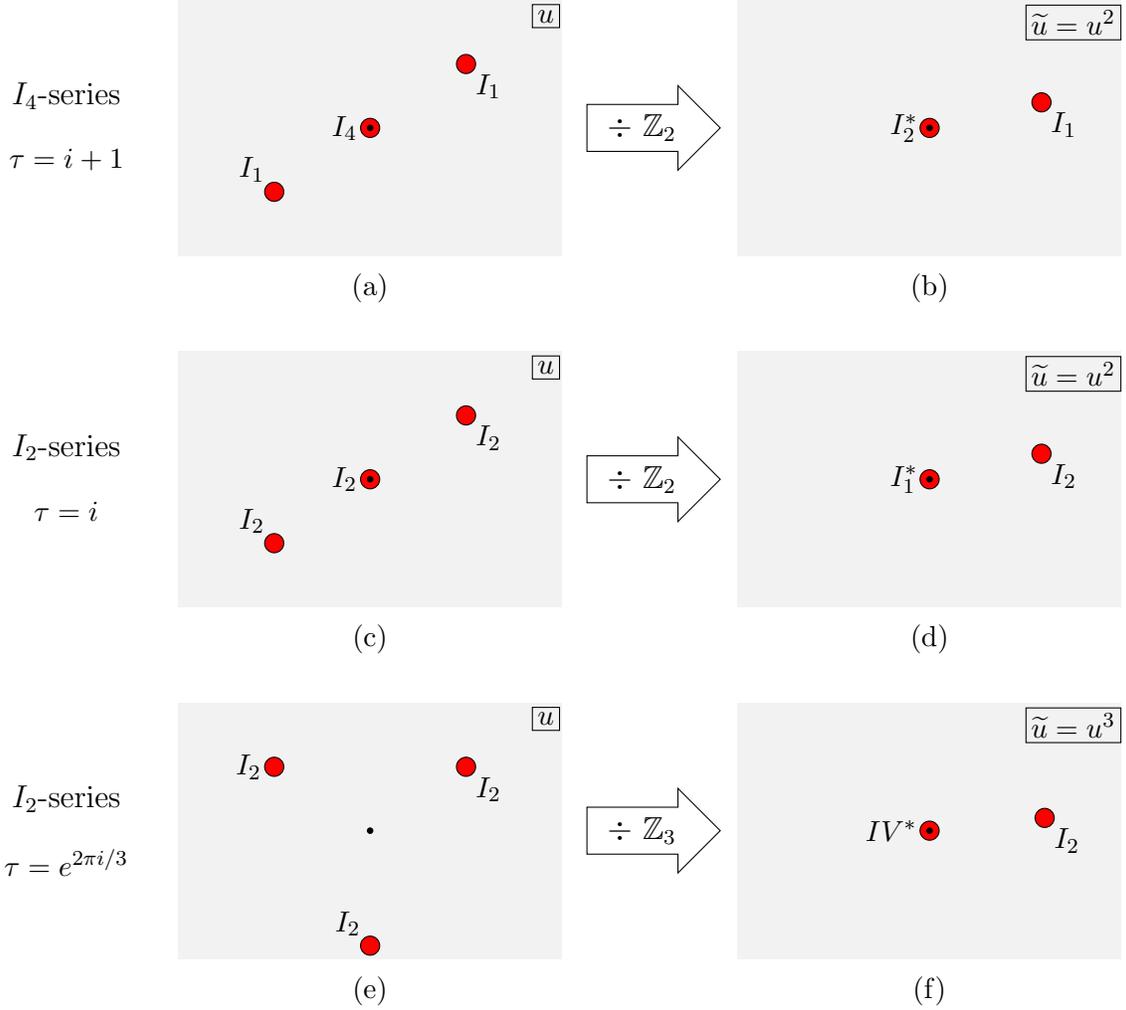
\begin{figure}[tbp]
\centering
\begin{tikzpicture}
\begin{scope}[scale=.85]
\begin{scope}[yshift=0cm]
\begin{scope}[xshift=-4.75cm]
\node at (0,.5) {\large $I_4$-series};
\node at (0,-.5) {$\t=i+1$};
\end{scope}
\begin{scope}[xshift=0cm]
\fill[color=black!05] (-3,-2) rectangle (3,2);
\node at (3-.25,2-.25) [rectangle,draw,inner sep=2pt,align=center]{$u$};
\node[R] at (0,0) {};
\node[R] at (+1.5,+1) {};
\node[R] at (-1.5,-1) {};
\node[bl] at (0,0) {};
\node at (-.4,0) {$I_4$};
\node at (+1.5+.35,+1-.35) {$I_1$};
\node at (-1.5-.35,-1+.35) {$I_1$};
\node at (0,-2.5) {(a)};
\end{scope}
\begin{scope}[xshift=4.25cm]
\node[single arrow, draw, black] at (0,0) {{\large $\ \div\ \Z_2\ $}};
\end{scope}
\begin{scope}[xshift=8.75cm]
\fill[color=black!05] (-3,-2) rectangle (3,2);
\node at (3-.75,2-.35) [rectangle,draw,inner sep=2pt,align=center]{$\til u = u^2$};
\node[R] at (0,0) {};
\node[R] at (+1.75,+.4) {};
\node[bl] at (0,0) {};
\node at (-.4,0) {$I_2^*$};
\node at (+1.75+.35,+.4-.35) {$I_1$};
\node at (0,-2.5) {(b)};
\end{scope}
\end{scope}
\begin{scope}[yshift=-5.5cm]
\begin{scope}[xshift=-4.75cm]
\node at (0,.5) {\large $I_2$-series};
\node at (0,-.5) {$\t=i$};
\end{scope}
\begin{scope}[xshift=0cm]
\fill[color=black!05] (-3,-2) rectangle (3,2);
\node at (3-.25,2-.25) [rectangle,draw,inner sep=2pt,align=center]{$u$};
\node[R] at (0,0) {};
\node[R] at (+1.5,+1) {};
\node[R] at (-1.5,-1) {};
\node[bl] at (0,0) {};
\node at (-.4,0) {$I_2$};
\node at (+1.5+.35,+1-.35) {$I_2$};
\node at (-1.5-.35,-1+.35) {$I_2$};
\node at (0,-2.5) {(c)};
\end{scope}
\begin{scope}[xshift=4.25cm]
\node[single arrow, draw, black] at (0,0) {{\large $\ \div\ \Z_2\ $}};
\end{scope}
\begin{scope}[xshift=8.75cm]
\fill[color=black!05] (-3,-2) rectangle (3,2);
\node at (3-.75,2-.35) [rectangle,draw,inner sep=2pt,align=center]{$\til u = u^2$};
\node[R] at (0,0) {};
\node[R] at (+1.75,+.4) {};
\node[bl] at (0,0) {};
\node at (-.4,0) {$I_1^*$};
\node at (+1.75+.35,+.4-.35) {$I_2$};
\node at (0,-2.5) {(d)};
\end{scope}
\end{scope}
\begin{scope}[yshift=-11cm]
\begin{scope}[xshift=-4.75cm]
\node at (0,.5) {\large $I_2$-series};
\node at (0,-.5) {$\t=e^{2\pi i/3}$};
\end{scope}
\begin{scope}[xshift=0cm]
\fill[color=black!05] (-3,-2) rectangle (3,2);
\node at (3-.25,2-.25) [rectangle,draw,inner sep=2pt,align=center]{$u$};
\node[R] at (-1.5,+1) {};
\node[R] at (+1.5,+1) {};
\node[R] at (0,-1.8) {};
\node[bl] at (0,0) {};
\node at (-1.5-.4,+1) {$I_2$};
\node at (+1.5+.35,+1-.35) {$I_2$};
\node at (-.35,-1.8+.35) {$I_2$};
\node at (0,-2.5) {(e)};
\end{scope}
\begin{scope}[xshift=4.25cm]
\node[single arrow, draw, black] at (0,0) {{\large $\ \div\ \Z_3\ $}};
\end{scope}
\begin{scope}[xshift=8.75cm]
\fill[color=black!05] (-3,-2) rectangle (3,2);
\node at (3-.75,2-.35) [rectangle,draw,inner sep=2pt,align=center]{$\til u = u^3$};
\node[R] at (0,0) {};
\node[R] at (+1.8,+.2) {};
\node[bl] at (0,0) {};
\node at (-.6,0) {$IV^*$};
\node at (+1.8+.35,+.2-.35) {$I_2$};
\node at (0,-2.5) {(f)};
\end{scope}
\end{scope}
\end{scope}
\end{tikzpicture}
\caption{The figures on the left are CB geometries of the $I_4$-series and $I_2$-series deformed $I_0^*$ geometry for special values of $\t$, and their daughter geometries are on the right.  The red circles mark the positions of the singularities, and the black dot marks the origin.  }
\label{fig3}
\end{figure}

From the prespective of their deformed CB geometries, the existence or abscence of these $\Z_2$ and $\Z_3$ symmetries for the $I_4$-series and $I_2$-series theories becomes almost obvious.  The $I_4$-series singularity splits into three as $I_0^* \to \{ I_4, I_1, I_1\}$ whose positions are governed by the zeros of the discriminant of its SW curve (constructed in \cite{paper2} and stated in appendix \ref{A2} below).  For general values of its marginal coupling and mass deformation parameters, $\t$ and $M$, these three singularities are at unsymmetrical positions on the CB.  But for $\t=i+1$ they exhibit a $\Z_2$ symmetry for arbitrary $M$, as shown in figure \ref{fig3}(a).  Upon gauging this $\Z_2$, the undeformable $I_4$ singularity at the origin becomes a frozen $I_2^*$ according to table \ref{Geometries}, while the two symmetrically placed $I_1$'s are identified.  Thus the daughter theory must have the CB geometry with deformation pattern $III^*\to\{I_2^*,I_1\}$, shown in figure \ref{fig3}(b).  This is derived algebraically from the form of the SW curve in appendix \ref{A2}.
It is also clear that there is no value of $\t$ in the parent $I_0^*$ theory where there is a $\Z_3$-symmetric CB, simply because two of the singularities are $I_1$'s while the third is an $I_4$.  

On the other hand, the $I_2$-series singularity splits as $I_0^* \to \{ I_2, I_2, I_2\}$ according to its SW curve \cite{sw2, paper2}, reviewed in appendix \ref{A2}.  For $\t=i$ the geometry is $\Z_2$-symmetric, figure \ref{fig3}(c). Upon gauging the $\Z_2$ the undeformable $I_2$ at the origin becomes a frozen $I_0^*$ according to table \ref{Geometries}, while the two symmetrically placed $I_2$'s are identified.  Thus the daughter theory has CB a geometry with deformation pattern $III^*\to\{I_0^*,I_2\}$, shown in figure \ref{fig3}(d).  But now, since all the singularities are $I_2$'s there can exist a $\Z_3$-symmetric configuration, which occurs at $\t=e^{2\pi i/3}$, figure \ref{fig3}(e).  Gauging this $\Z_3$ makes the free $I_0$ theory at the origin into a frozen $IV^*$ singularity according to table \ref{Geometries2}, while the three symmetrically placed $I_2$'s are identified.  Thus the daughter theory has CB a geometry with deformation pattern $III^*\to\{IV^*, I_2\}$, shown in figure \ref{fig3}(e). Again see appendix \ref{A2} for the explicit derivation of these facts from the SW curve.

The existence of a $\Z_3$ discrete gauging of the $I_0^*\to\{I_2^3\}$ has striking implications, as there is no $\Z_3$ invariant orbifold point in the fundamental domain of the (standard) $\cN=4$ $\SU(2)$ SYM theory, as argued in \cite{Aharony:2013hda}.  Yet the $\Z_3$ discrete gauging of the $I_0^*\to\{I_2^3\}$ passes all our non-trivial consistency checks, which suggests that this second geometry should be associated to a different $\SU(2)$ $\cN=4$ theory, likely with a different spectrum of line operators than those presented in \cite{Aharony:2013hda}.  The $I_0^*\to\{I_1^2,I_4\}$ geometry has instead all the properties of the standard $\cN=4$ $\SU(2)$ theory described in \cite{Aharony:2013hda}. We will elaborate further on this in an upcoming paper \cite{paper2c}.

\section{Higgs branches}\label{sec:HB}

In this section we present a detailed analysis on how gauging a discrete symmetry acts on the Higgs branch chiral ring. Working out in detail one particular example will be illuminating to understand the action of the discrete gauging procedure on local operators. Also the results we find are entirely consistent with the Higgs branch constraints which can be extracted from the $(c,k_F)$ central charge data, as explained in \cite{b+13, Lemos:2015orc}. As we will explain shortly, the way things work out is highly non-trivial and to our knowledge such intricate Higgs branch construction was not seen before. 

For a detailed geometrical and algebraic description of Higgs branches of $\cN=2$ SCFTs, as well as as a careful description of the notation used in this section, we refer to \cite{Dolan:2002zh,Beem:2014zpa,Argyres:2016xmc}. 

\subsection{An example in detail}

The most convenient example to study is 
\begin{align}\nonumber
\ [I_0^*,D_4] \ 
\xrightarrow{\text{$\Z_2$ gauging}}
\ [III^*,B_3],
\end{align}
which has the advantage that we are able to carry out calculations explicitly.  We start by reminding the reader about the structure of the Higgs branch of the $[I_0^*,D_4]$ theory.  Recall that $D_4$ and $B_3$ are the Dynkin notation for the $\SO(8)$ and $\SO(7)$ Lie algebras, respectively.  In this section we will use the $(D_4,B_3)$ and the more familiar $(\SO(8),\SO(7))$ notations interchangeably.  

\paragraph{[\emph{I}$_{\bf 0}^*$,\emph{D}$_{\bf4}$] Higgs branch.} 

This theory has a well known Lagrangian description as the $\cN=2$ $\SU(2)$ theory with 4 hypermultiplets in the fundamental representation $\bf 2$. Because the $\bf 2$ is a pseudo-real representation the chiral multiplets can be re-organized counting the 8 half-hypers instead which transform under the $\SO(8)$ flavor symmetry. We can then denote the field content by $Q^a_i$ where $i=1,...,8$ is a flavor index and labels the half-hypers, while $a=1,2$ is an $\SU(2)$ index. $Q_i^a$ transforms in the ${\bf 8}_v$ of the flavor $\SO(8)$. The Higgs branch chiral ring is generated by a single dimension two operator transforming in the adjoint of $\SO(8)$:
\beq\label{MesonD4}
M_{[ij]}:=Q_{ai}Q^a_j
\eeq
where the $\SU(2)$ index is lowered with the invariant $\epsilon_{ab}$ tensor and the square brackets indicate that $M$ is antisymmetric in $i$ and $j$. \eqref{MesonD4} is the usual meson operator which is identified with the scalar primary of the $\cBh_1$ of the $D_4$ theory which, following \cite{Argyres:1996eh} will be labeled by $q_1$, the 1 labeling the $\SU(2)_R$ ``spin'' of the operator:
\ytableausetup{boxsize=1.2mm}
\beq
M_{[ij]}= q_1^{{\bf 28}} 
\qquad\in \cBh_1 \qquad \text{with}\qquad R=1,\ r=0,
\eeq 
the superscript of the $q_1$ operator indicates its $\SO(8)$ representation.

As extensively explained in the literature (see for example \cite{Beem:2014zpa,Argyres:2016xmc}) the Higgs branch chiral ring is generically not freely generated and the $q_1$'s satisfy non-trivial relations. After imposing the $F$ and $D$ term condition, these relations for the $D_4$ theory can be written as follows:
\beq\label{D4relations}
q_1^{{\bf28}} q_1^{{\bf28}} \sim q_2^{{\bf300}}.
\eeq
It is helpful to recall the representation theory of the symmetrized tensor product ($\otimes_S$) of adjoint representations of $\SO(8)$:
\ytableausetup{boxsize=1.5mm}
\beq
{\bf28}\otimes_S{\bf28}={\bf300}\oplus\left[{\bf35}_v\oplus{\bf35}_c\oplus{\bf35}_s\oplus{\bf1}\right]
\eeq
Relations in \eqref{D4relations} imply then that the $q_2$'s, scalar primaries of the $\cBh_2$ operators, transforming in the representation in the square bracket above, should not appear in the OPE of the $\cBh_1$. As discussed in detail in \cite{b+13, Lemos:2015orc}, these OPE coefficients can be set to zero {\it if and only if} the flavor central charge $k_F$ and the $c$ anomaly coefficient saturate certain flavor algebra dependent bounds which for $D_4$ give $k_F=4$ and $12c=14$.  These are precisely the values of the $(c,k_F)$ central charges of the $D_4$ theory and thus \eqref{D4relations} follows.  The $[\cdot]=0$ relations generate the $D_4$ {\it Joseph ideal} and the $q_1$'s satisfying such relations describe the {\it minimal nilpotent orbit} of $D_4$ which in the physics literature is also known as the {\it centered one instanton moduli space}; see for example \cite{Gaiotto:2008nz, b+13}.  As pointed out in the beautiful work \cite{b+13,  Lemos:2015orc}, only SCFTs with a very restricted set of flavor algebras, namely $A_1$, $A_2$, $D_4$, $G_2$, $F_4$, $E_6$, $E_7$, $E_8$ can have a one instanton moduli space Higgs branch.

The relation in \eqref{D4relations} completely characterizes the Higgs branch of the theory.

\paragraph{$[III^*,B_3]$ Higgs branch.} 

To start recall that this theory is obtained from the $[I_0^*,D_4]$ by modding out by a chosen, yet arbitrary, $\Z_2$ subgroup of the outer automorphism group of $D_4$, $\Out(D_4) \simeq S_3$. Because the generators of the $[I_0^*,D_4]$ Higgs branch chiral ring transform non-trivially under the flavor group, and thus under the gauged $\Z_2$, we expect the Higgs branch chiral ring of $[III^*,B_3]$ to differ from that of the $[I_0^*,D_4]$ Higgs branch.  The relations satisfied by the $q_1^{B_3}$ can be explicitly computed from the chiral ring relations of the $[I_0^*,D_4]$ theory \eqref{D4relations}.

Modding out by the outer $\Z_2$ picks an $\SO(7)$ Lie algebra within the original $\SO(8)$.  Under the $\SO(7)$ the $q_1$ decompose as
\ytableausetup{boxsize=1.2mm}
\beq\label{B3deco}
q_1^{{\bf28}}\quad\xrightarrow{\SO(7)}\quad \tilde{q}_1^{{\bf21}}\oplus \tilde{q}_1^{{\bf7}}
\eeq
where we label the scalar primaries of the $\cBh_1$ operator for the $B_3$ theory as $\tilde{q}_1$.

The transformation of the $\tilde{q}_1$ under the $\Z_2$ can be obtained by choosing an explicit form for the $\Z_2$ action on the half-hypers of the $[I_0^*,D_4]$ theory:
\begin{align}
\begin{array}{c}
Q_I^a\\\label{Z2action}
Q_8^a
\end{array}\qquad\xrightarrow{\ \Z_2}\qquad \begin{array}{c}
Q_I^a\\
-Q_8^a
\end{array}\qquad \text{with}\ I=1,...,7.
\end{align}
From \eqref{MesonD4}, it is straightforward to identify $\tilde{q}_1^{{\bf21}}:=M_{[IJ]}$, $I,J=1,...,7$ and $q_1^{{\bf7}}:=M_{[I8]}$. From \eqref{Z2action} follows
\begin{align}
\begin{array}{c}
\tilde{q}_1^{{\bf21}}\\\label{Z2actionq}
\tilde{q}_1^{{\bf7}}
\end{array}\qquad\xrightarrow{\ \Z_2}\qquad \begin{array}{c}
\tilde{q}_1^{{\bf21}}\\
-\tilde{q}_1^{{\bf7}}
\end{array}
\end{align}
Thus at the level of the $\cBh_1$ operator, gauging the $\Z_2$ eliminates all but the scalar primary which transforms under the adjoint of the $\SO(7)$ flavor group, as expected.  Nevertheless, the $\tilde{q}_1^{{\bf7}}$ are eliminated from the theory altogether:  they ``make it back" in the theory at the level of the $\cBh_2$ as we will explain now.

From \eqref{Z2actionq} it trivially follows that we can form a $\Z_2$-invariant tensor by pairing up two $\tilde{q}_1^{\bf7}$ operators.  So any operator obtained from $\tilde{q}_1^{{\bf7}}\tilde{q}_1^{{\bf7}}$ should be kept in the theory along with the operators obtained from $\tilde{q}_1^{{\bf21}}\tilde{q}_1^{{\bf21}}$. Operators obtained from $\tilde{q}_1^{\bf7}\tilde{q}_1^{\bf21}$ are instead odd under the $\Z_2$ and should be eliminated after gauging.  As explained above, there are also non-trivial relations inherited by the relations satisfied by the $q_1$'s in \eqref{D4relations}.  Let's first summarize the relevant $B_3$ group theory for the case at hand (we write in red the representations which needs to be crossed out by gauging the $\Z_2$):
\ytableausetup{boxsize=1.5mm}%
\begin{align}\label{axsaB3}
{\bf21} \otimes_S {\bf21}&= {\bf168} \oplus\left[{\bf27}\oplus{\bf35}\oplus{\bf1}\right],
\\ \label{axfB3}
{\bf7} \otimes_S{\bf7} &= {\bf27} \oplus{\bf1},\\
\red{{\bf21} \otimes{\bf7}} &\red{= {\bf105} \oplus{\bf35}\oplus {\bf7}}.
\end{align}
Again the relations $[\cdot]=0$ describe the $B_3$ Joseph ideal. The next step is write the relations satisfied by the $q_1$ operators in the $D_4$ case in terms of $\SO(7)$ representations: 
\begin{align}\label{D4B30}
[\cdot]_{D_4}\quad&\to\quad2({\bf35})\oplus{\bf27}\oplus{\bf7}\oplus{\bf1}=0
\\ \label{D4B3n0}
{\bf300}_{D_4}\quad&\to\quad{\bf168}\oplus{\bf105}\oplus{\bf27}\neq0
\end{align}
Here ``$\neq0$" means ``lack of any relation".
\eqref{D4B3n0} implies the following (lack of) relations on the OPE of the $\tilde{q}_1$:
\beq
\tilde{q}_1^{\bf21}\tilde{q}_1^{\bf21}|_{\bf168}\neq0,\qquad\qquad\tilde{q}_1^{\bf7}\tilde{q}_1^{\bf7}|_{\bf27}\neq0\qquad\text{and}\qquad\tilde{q}_1^{\bf7}\tilde{q}_1^{\bf21}|_{\bf105}\neq0 .
\eeq
The first two operators are invariant under the $\Z_2$ and thus we expect the $[III^*,B_3]$ theory to have both a $\tilde{q}_2^{\bf168}$ and a $\tilde{q}_2^{\bf27}$ in the spectrum.  The third operator is projected out by the $\Z_2$.

Let us now analyze the relations inherited from \eqref{D4B30}.
At first sight one might think that the $D_4$ Joseph ideal relation implies that the $\tilde{q}_1^{{\bf21}}$ describe a $B_3$ minimal nilpotent orbit. Yet from the conformal bootstrap analysis, as previously mentioned, no $B_3$ theory can have a minimal nilpotent orbit Higgs branch component. To resolve this conundrum we need to analyze more carefully the structure of the decomposition of the $D_4$ Joseph ideal into $\SO(7)$ representations. We will find that in fact one of the $D_4$ Joseph ideal relations implies that $\tilde{q}_1^{{\bf21}}\tilde{q}_1^{{\bf21}}|_{{\bf27}}=\tilde{q}_1^{{\bf7}}\tilde{q}_1^{{\bf7}}|_{{\bf27}}$ and the corresponding term in the OPE does not vanish, thus providing a perfectly consistent Higgs branch chiral ring.  In this way the $\cBh_2^{\bf27}$ ``makes it back'' into the $\cBh^{\bf21}_1\cBh^{\bf21}_1$ OPE after imposing the $D_4$ Joseph ideal relations.

To show that this is the case, let's first explicitly write down $\tilde{q}_1^{{\bf21}}\tilde{q}_1^{{\bf21}}|_{{\bf27}}$ and $\tilde{q}_1^{{\bf7}}\tilde{q}_1^{{\bf7}}|_{{\bf27}}$ in terms of the meson operator \eqref{MesonD4}:
\begin{align}
\tilde{q}_1^{{\bf21}}\tilde{q}_1^{{\bf21}}|_{{\bf27}}&\equiv M_{(I}^{\phantom{ii}j}M_{jK)}\\
\tilde{q}_1^{{\bf7}}\tilde{q}_1^{{\bf7}}|_{{\bf27}}&\equiv M_{8(i}M_{j)8}
\end{align}
Indices are raised using the $\delta^{ik}$ $\SO(8)$ invariant tensor, the antisymmetrization of the $M$ indices is implicit while we explicitly write the symmetrization of $I$ and $K$. Capital indices only go up to 7. Let's now work out the $\SO(8)\to\SO(7)$ decomposition of the relevant representation that in this case is the ${\bf 35}_c$.  The ${\bf35}$, in terms of the meson operator can be written as
\ytableausetup{boxsize=1.7mm}
\beq\label{35repr}
{\bf35}_c\equiv M_{(i}^{\phantom i j}M_{jk)}-M^{ij}M_{ij},\qquad i,j,k,l=1,..,8.
\eeq
The representation above corresponds to the Young tableau $\yd{2}$\ . Using $[\cdot]_{D_4}|_{\bf1}=M^{ij}M_{ij}=0$ we can write \eqref{35repr} in terms of the chosen $\SO(7)$ embedding as follows:
\beq
{\bf35}_c\quad\xrightarrow{\SO(7)}\quad
\left\{
\begin{array}{l}
M_I^{\phantom i j}M_{jK}\\
M_I^{\phantom i j}M_{j8}\\
M_8^{\phantom i j}M_{j8}\\
\end{array}
\quad\text{transforming\ as}\quad
\begin{array}{l}
{\bf27}\\
{\bf7}\\
{\bf1}
\end{array}
\right.
\eeq 
Splitting the sum over $j$, and setting the ${\bf 27}$ to zero we obtain:
\beq
M_{(I}^{\phantom i J}M_{JK)}=-M_I^{\phantom i 8}M_{8K}\qquad\text{or}\qquad \tilde{q}_1^{{\bf21}}\tilde{q}_1^{{\bf21}}|_{{\bf27}}\propto\tilde{q}_1^{{\bf7}}\tilde{q}_1^{{\bf7}}|_{{\bf27}}
\eeq
from which we can obtain the following $[III^*,B_3]$ Higgs branch chiral ring relations:
\beq
\tilde{q}_1^{\bf21}\tilde{q}_1^{\bf21}\sim \tilde{q}_2^{{\bf 168}}+\tilde{q}_2^{\bf 27}
\eeq
which, if compared with \eqref{axsaB3}, imply that $\tilde{q}_1^{\bf21}\tilde{q}_1^{\bf21}|_{\bf35}=\tilde{q}_1^{\bf21}\tilde{q}_1^{\bf21}|_{\bf1}=0$. 

These constraints are exactly what we expect from the values of the $(k_F,c)$ central charges of the $[III^*,B_3]$ theory, as we now explain.  The values of the $(k_F,c)$ charges dictate the vanishing of certain OPE coefficients corresponding to $\cBh_2$ operators transforming in specific representaitons in the $adj\otimes_S adj$ \cite{b+13, Beem:2014zpa, Lemos:2015orc}. Since the $[III^*,B_3]$ theory is obtained from gauging a discrete flavor group of the $[I_0^*,D_4]$, the central charges of the former are equal to the ones of the latter which are known to be $(k_F=4,12c=14)$. $k_F=4$ for a $B_3$ theory saturates a bound corresponding to setting to zero precisely the OPE coefficient for the ${\bf35}$ (see table 3 of \cite{b+13}). Furthermore the OPE coefficient corresponding to the singlet representation, is zero only when the following, Lie algebra dependent, bound is satisfied \cite{Beem:2014rza}:
\beq\label{singCond}
\frac{1}{k}=\frac{12 c+\rm{dim}_G}{24c \ h^{\vee}}
\eeq
where dim$_G$ and $h^{\vee}$ are the dimension and the dual Coxeter number of the flavor Lie algebra. From table \ref{Tab:groupinfo} we can extract these values for the $B_3$ case and check that \ref{singCond} is satisfied precisely for $(k_F=4,12c=14)$. This observation concludes the presentation of a beautifully consistent picture for the Higgs branch chiral rings of discretely gauged theories. As we are going to describe next, a very similar story applies to all the other theories with gauged discrete groups. 

\begin{table}[t!] 
\centering
\begin{tabular}{lll||lll}
\hline \hline
$G$ 			&	$h^{\vee}$		& $\mathrm{dim}_{G}$		&	$G$ 			&	$h^{\vee}$		& $\mathrm{dim}_{G}$		\\[0.3ex]
\hline 
$\SU(N)\ (A_{N-1})$		&	$N$				& $N^2-1$				&	$F_4$		&	$9$				& $52$					\\[0.3ex]
$SO(N)\ (B_{(N-1)/2}\ \&\ D_{N/2})$		& 	$N-2$			& $\frac{N(N-1)}{2}$		&	$E_6$~~~		&	$12$	~~~			& $78$~~~~				\\[0.3ex]
$\Sp(2N)\ (C_N)$	&	$N+1$			& $N(2N+1)$				&	$E_7$		&	$18$				& $133$					\\[0.3ex]
$G_2$		&	$4$				& $14$					&	$E_8$		&	$30$				& $248$					\\[0.3ex]
\hline
\end{tabular} 
\caption{Dual Coxeter number and dimension of the adjoint representation for the simple Lie groups.
\label{Tab:groupinfo}}
\end{table} 

\subsection{Higgs branches for theories with disconnected gauge groups}

In this subsection we report the Higgs branch chiral rings of the remaining theories in table \ref{tab1}. For most of these theories no lagrangian description is available and it is not possible to perform a detailed analysis like the one described above. The results reported below are obtained using representation theory and asking for consistency with respect to the central charges values. The Higgs branch of the $[IV^*,A_2]$ and $[III^*,A_1]$ are equivalent to the Higgs branch of the $[IV,A_2]$ and $[III,A_1]$, that is they span the minimal nilpotent orbit of $A_2$ and $A_1$ respectively. In fact in these cases the discrete gauging does not carry any action on the flavor symmetry algebra.

\paragraph{$[II^*,F_4]$ Higgs branch.} 

This theory is obtained from the $[IV^*,E_6]$. Gauging the Out$(E_6)=\Z_2$ we obtain a $F_4$ theory. Under the $\Z_2$ the $\cBh_1$ operator of the $[IV^*,E_6]$ decomposes into an even part $\cBh^{\bf\ 52}_1$ which has the proper flavor transformation to be identified with the $\cBh_1$ for the $F_4$ theory, and a $\cBh_1^{\bf\ 26}$ which is odd and is eliminated from the theory.  At the level of the $\cBh_2$ operator, we need to study the reduction of the $E_6$ Joseph ideal relations which work as follows:
\beq
\begin{array}{ccccc}
\bf{78}\otimes_S{\bf78}&=&{\bf2430}&\oplus&[{\bf650}\oplus{\bf1}]\\
\bigg\downarrow &&\ \ \bigg \downarrow{F_4 }&&\bigg\downarrow\\
{\bf52}\otimes_S{\bf52}&=& {\bf1053'}&\oplus &{\bf324}\oplus{\bf1}\\
{\bf26}\otimes_S{\bf26}&=& {\bf324}&\oplus& {\bf26}\oplus{\bf1}\\
\red{{\bf26}\otimes{\bf52}}&\red{=}& \red{{\bf1053}}&\red{\oplus}& \red{{\bf273}\oplus{\bf26}}
\end{array}
\eeq
No operator associated to the representations in red should appear in the theory as those correspond to operators obtained from a $q_1^{\bf52}q_1^{\bf26}$ product which are all $\Z_2$ odd. An argument similar to the one described above can be used to guess the following Higgs chiral ring relations for the $[II^*,F_4]$ theory
\beq
q_1^{\bf52}q_1^{\bf52}\sim q_2^{\bf1053'}+q_2^{\bf324}
\eeq
which then only implies the constrain $q_2|_{\bf1}=0$. This is compatible with the fact that plugging the values of $(c,k_F)_{E_6}$ and the (dim$_G$, $h^{\vee})_{F_4}$ in the \eqref{singCond}, the central charges of the obtained $[II^*,F_4]$ saturate the appropriate bound for the vanishing of the OPE coefficient associated to the singlet channel.

\paragraph{$[II^*,G_2]$ Higgs branch.} 

This theory is instead obtained from the $[I_0^*,D_4]$ and the $G_2$ flavor symmetry is obtained by gauging by a $\Z_3$ subgroup of the $S_3$ outer automorphism group of $D_4$. This case is slightly more involved than the previous one because the $\cBh_1$ operator of the $D_4$ theory decomposes in three components: $\cBh_1^{\bf14}$ which is invariant under the $\Z_3$ and needs to be identified with the $\cBh_1$ of the $[II^*,G_2]$, and two $\cBh^{\bf7}_1$'s, with $\Z_3$ charges $\pm 1$ (mod 3) which we will denote by $\cBh_1^{{\bf7}\pm}$. From those $\cBh_1$ operators we can form, at the quadratic level, combinations with $\Z_3$ charge 0, $+1$, and $-1$. The representations of those operators will be indicated in the table below in black, green and blue respectively:
\beq
\begin{array}{ccccc}
\bf{28}\otimes_S{\bf28}&=&{\bf300}&\oplus&[{\bf35}\oplus{\bf35}_v\oplus{\bf35}_s\oplus{\bf35}_c\oplus{\bf1}]\\
\bigg\downarrow &&\ \ \bigg \downarrow{G_2 }&&\bigg\downarrow\\
{\bf14}\otimes_S{\bf14}&=& {\bf77}&\oplus &{\bf27}\oplus{\bf1}\\
{\bf7^+}\otimes{\bf7^-}&=& {\bf27}&\oplus& {\bf1}\\
\green{{\bf7^+}\otimes{\bf14}}&\green{=}& \green{{\bf64}}&\green{\oplus}& \green{{\bf27}\oplus{\bf7}}\\
\green{{\bf7^-}\otimes_S{\bf7^-}}&\green{=}& \green{{\bf27}}&\green{\oplus}& \green{{\bf1}}\\
\blue{{\bf7^+}\otimes_S{\bf7^+}}&\blue{=}& \blue{{\bf27}}&\blue{\oplus}& \blue{{\bf1}}\\
\blue{{\bf7^-}\otimes{\bf14}}&\blue{=}& \blue{{\bf64}}&\blue{\oplus}& \blue{{\bf27}\oplus{\bf7}}
\end{array}
\eeq
In this case the $\Z_3$ modding gets rid of all the operators associated to the representations in green and blue. Following the same argument as above, we can guess the following Higgs chiral ring relations for the $[II^*,G_2]$ theory
\beq
q_1^{\bf14}q_1^{\bf14}\sim q_2^{\bf77}+q_2^{\bf27}
\eeq
which again implies the constraint $q_2|_{\bf1}=0$. This is result is remarkable as \eqref{singCond}, with $(c,k_F)_{D_4}$, can be saturate not just with (dim$_G$, $h^{\vee})_{D_4/B_3}$, as already shown above, but also by (dim$_G$, $h^{\vee})_{G_2}$ from table \ref{Tab:groupinfo} giving again a beautifully consistent picture.

\begin{center}
---\ \ $*$\ \ ---
\end{center}

The structure of the Higgs branch of the $[IV,A_2]\xrightarrow{\text{$\Z_6$ gauging}}[II^*,A_1]$ follows a similar derivation. The Higgs branch of $[IV,A_2]$ is also the centered one-instanton moduli space of $A_2$ and the one of $[II^*,A_1]$ is obtained by carefully going through the representation theory analysis. The case of the $[II^*,C_2]$ is complicated by the fact that the Higgs branch of the parent $[IV^*,C_2U_1]$ theory is not the minimal nilpotent orbit of $C_2$ and it is in fact a 8 complex dimensional variety. In \cite{Argyres:2016xmc} we observed that $C_2$ has a unique, special, nilpotent orbit of complex dimension 8 and it thus tempting to identify the Higgs branch of $[IV^*,C_2]$ with it. Yet we don't know of a nice parametrization of this orbit like in the minimal case and thus the arguments above do not apply straightforwardly to this case.

\section{Central charges}\label{sec:CC}

In \cite{Argyres:2016xmc}, generalizing the beautiful work of \cite{st08}, we were able to derive a series of formulae to compute the $a$ and $c$ central charges from the deformation pattern of a given SCFT.  As explained above and in more detail in \cite{paper1,paper2,Argyres:2016xmc}, turning on relevant parameters deforms the CB singularity associated to the SCFT into lesser ones. When all available relevant deformations are switched on the SCFT singularity is maximally split into frozen or undeformable singularities. These singularities form the deformation pattern of the initial singularity and they are identified with particular IR free lagrangian theories whose $a$ and $c$ central charges are known. Calling $a_i$ and $c_i$ the known central charges of the $i$-th singularity, the central charges of the initial SCFT are \cite{Argyres:2016xmc}:
\begin{align}
24 a &= 5 + h + 6 (\D-1) + \D \sum_{i=1}^Z \frac{12 c_i -2-h_i}{\D_i},
\label{ac-formula1}\\\label{ac-formula2}
12 c &= 2 + h + \D \sum_{i=1}^Z \frac{12 c_i -2-h_i}{\D_i} .
\end{align}
where $h$ is the quaternionic dimension of the mixed branch of the SCFT while the sum and the values for $c_i$, $\Delta_i$, and $h_i$, refer to the singularities in the deformation pattern.  Adding \eqref{ac-formula1} and \eqref{ac-formula2} we can obtain a relation between the $(a,c)$ central charges and $\Delta$, the scaling dimension of the CB vev, derived first by Shapere and Tachikawa in \cite{st08} 
\beq\label{acu}
2a-c=\frac{2\Delta-1}{4}.
\eeq

As was noted first in \cite{Aharony:2016kai}, this relation is clearly violated by SCFTs obtained by gauging discrete symmetries, since this operation does not change the central charges, but does change the dimension of the CB parameter.
However, a simple modification of \eqref{ac-formula2} gives the correct results:
\begin{align}\label{ac-formulaDG}
24 a_{\Z_r} &= 5 + h + 6 \left(\frac{\D}{r}-1\right) + \D \sum_{i=1}^Z \frac{12 c_i -2-h_i}{\D_i}.
\end{align}
Here $r$ refers to $\Z_r$ action of the discretely-gauged group on the parent CB, while all the other parameters describe the corresponding quantities in the daughter theory.  Equation \eqref{ac-formula2} remains unchanged, but again with the parameters referring to the daughter theory.

While \eqref{ac-formulaDG} works empirically, it seems challenging to derive if from the twisted CB partition function argument that gave \eqref{ac-formula1} and \eqref{ac-formula2}.  
The reason is that it involves the quantity ``$r$" which refers to a property of the parent theory, and not obviously intrinsic to the low energy effective action of the daughter theory.  Note that, following the arguments of \cite{Gaiotto:2014kfa}, $\Z_r$ is expected to be related to the 2-form global symmetry of the daughter theory.  But it is less than clear how the associated surface operators can contribute to the CB twisted partition function to give \eqref{ac-formulaDG}.


\section{Conclusion and open questions}

We have presented a systematic study of $\cN=2$ SUSY preserving gaugings of discrete global symmetry in the context of four dimensional rank-1 $\cN=2$ field theories.  We recast the discussion of gauging a discrete symmetry in a free $\cN=4$ field theory in \cite{gr1512, Aharony:2016kai} in an $\cN=2$ language.  This laid the groundwork for a systematic study of allowed $\Z_k$ discrete gaugings of general $U(1)$ $\cN=2$ gauge theories.  In doing that and generalizing this construction to interacting non-lagrangian theories, we found that discrete gauging can be understood in a simple and beautiful way in terms of the CB geometry.  

We found:
\begin{itemize}
\item Only very special $\Z_k$ subgroups of $U(1)_R\times SL(2,\Z)\times\Out(F)$ preserve $\cN=2$ supersymmetry, generalizing the construction of \cite{gr1512}.

\item A $\Z_k\subset U(1)_R$ acts as a $\Z_{r=k/\D}$ on a parent theory with CB parameter of scaling dimension $\D$. The resulting daughter theory has a CB parametrized by $\til u=u^r$.

\item By gauging a $\Z_2$ symmetry of a $U(1)$ $\cN=2$ gauge theory with beta function $2n$, its $I_{2n}$ CB geometry is mapped to an $I_n^*$ geometry with unusual flavor symmetry, including frozen $I_n^*$ geometries for any $n$.  Similarly, gauging $\Z_3$, $\Z_4$ and $\Z_6$ symmetries of free $\cN=2$ $U(1)$ theories gives frozen $IV^*$, $III^*$ and $II^*$ CB geometries.

\item $\Z_k\subset SL(2,\Z)$ subgroups of the EM duality group are global symmetries for $\Z_k$-invariant values of the holomorphic gauge coupling $\t$.  This restricts the possible discrete groups which preserve $\cN=2$ supersymmetry of isolated SCFTs with a fixed $\t$.  Conversely, gauging such discrete symmetries of non-isolated SCFTs lifts their conformal manifold, fixing $\t$ to a particular value. 

\item Only a subset of the mass deformations of a parent theory with flavor symmetry $F$ preserves a $\Z_r$ symmetry of the CB.  Discretely gauging the $\Z_r$ allows only mass deformations which are fixed by a discrete subgroup $\G\subset \Out(F)$, so the daughter theory's flavor symmetry is $F'=F/\G$ for some action of $\G$ on $F$.  This determines the splitting of the daughter theory CB singularity under generic mass deformation, which is enough information \cite{paper2} to construct the CB geometry associated to the daughter theory.  Only a small set of symmetry algebras $F'$ are compatible with a given SW curve.  The existence of a solution of the form $F'=F/\G$ compatible with the constructed curve is a non-trivial consistency check.

\item We had previously constructed Seiberg-Witten geometries associated to deformation patterns containing frozen $I_0^*$ and $I_2^*$ singularities, but which had no known realization in terms of $\cN=2$ field theories.  They now have a beautifully consistent physical interpretation as discretely gauged versions of known theories.  Among these new theories are ones with $F_4$ and $G_2$ flavor symmetry algebra, as well as two new $\cN=3$ SCFTs.

\item The general formula to compute the $a$ central charge for a given SCFT knowing its deformation pattern \cite{Argyres:2016xmc} fails when applied to discretely gauged theories \cite{Aharony:2016kai}. This can be seen as a reflection of the fact that discretely gauged theories have the same central charges as their parent theories but different CB scaling dimension, $\D$, and so violate the relation between $a$, $c$ and $\D$ derived by Shapere and Tachikawa \cite{st08}.  We guessed a modified formula, \eqref{ac-formulaDG}, which works for computing the $a$ central charge for discretely gauged theories. 
\end{itemize}

While the picture presented in this paper is fairly complete and very consistent there are quite a few questions which remain open.  Apart from the obvious questions of whether string, S-class, or bootstrap methods can realize the rank-1 theories described here, and of the generalization of this story to higher-rank $\cN=2$ theories, here are some puzzles raised just within our rank-1 field theory analysis.
\begin{itemize}
\item As pointed out in section \ref{sec:CC}, we lack an intrinsic way to compute the central charges of the daughter theories, and instead have to refer back to their relation to ``the" parent theory to do so.  Following \cite{Gaiotto:2014kfa}, perhaps the missing intrinsic data is in the spectrum of surface operators of the daughter theories.

\item Two of the daughter theories in table \ref{tab1} appear twice:  the $[III^*,C_1]$ and $[III^*,U_1\rtimes\Z_2]$ theories appear as $\Z_2$ and $\green{\til \Z_2}$ daughters of both the $I_4$-series and $I_2$-series $[I_0^*,C_1\Chi_0]$ theories (they are in fact associated to two different deformation patterns, $III^*\to\{I_2^*,I_1\}$ and $III^*\to\{I_1^*,I_2\}$).  We have conjectured that these two parent theories are subtly different.  Is that also true of their daughters?

\item We have only discussed gauging $\Z_n$ symmetries, that is, discrete groups with a single generator.  We can also imagine gauging non-cyclic abelian discrete groups, e.g., $\Z_2\times\Z_2$.  One way of exploring this question with our method is to ask whether any of the daughter theories we found in table \ref{tab1} have further cyclic symmetries that could be consistently gauged.  The only possible non-free example of this is the $[IV^*,A_2]$ daughter of the $[IV,A_2\chi_{1/2}]$ parent, which has a further discrete symmetry which acts as a $\Z_2$ on its CB.  But gauging this symmetry seems to give the $[II^*,B_1]$ theory, the other daughter of the same $[IV,A_2\chi_{1/2}]$ parent, and so we find no obviously new theories in this way.  As in the previous question, it is possible that these two routes to constructing the $[II^*,B_1]$ daughter theory might be subtly different, e.g., their local operator algebras might be the same but their spectra of line and surface operators might differ as in \cite{Aharony:2013hda}.  This possibility might also apply to the free theories in table \ref{tab1} which can be reached by successive discrete $\Z_k$ gaugings.  Could these multiple versions of the free $\cN=3$ $\Z_k$ gaugings correspond to the multiple versions of these thoeries constructed via S-foldings \cite{gr1512, Aharony:2016kai}?  

\item Gauging non-abelian discrete groups is also interesting.  One might have expected the $[I_0^*,D_4\chi_0]$ theory to have a gaugable non-abelian discrete symmetry, since $\Out(D_4)\simeq S_3$.  However, its $\Z_2$ and $\Z_3$ subgroups combine with S-duality actions which fix different values of the marginal coupling, $\t$, and so cannot be realized simultaneously.  In general, the subgroups of $SL(2,\Z)$ which can be global symmetries are only cyclic groups.  At higher rank, there may be finite non-abelian subgroups of the $Sp(2r,\Z)$ EM-duality group which fix a given $r\times r$ matrix $\t_{ij}$ of low energy couplings.

\item Possibly the most puzzling part of our study is the role played by $\Inn(F)$, the group of inner automorphisms of the flavor symmetry algebra.  We have emphasized that discrete symmetries which act on the CB and commute with $\cN=2$ supersymmetry must involve the action of a subgroup $\Z_k\subset\Out(F)$ of the outer automorphism group of the flavor symmetry.  But this action is arbitrary up to the choice of an element $g\in\Inn(F)$, i.e., the $\Z_k$ generated by $\vf\in\Aut(F)$ and the $\til \Z_k$ generated by $\til\vf := g \vf g^{-1}$ may act differently on $F$ and so give different daughter flavor symmetries: $F/\Z_k \neq F/\til\Z_k$.  However, as explained in examples in sections \ref{U1matt} and \ref{sec3.3}, not all choices of $g\in\Inn(F)$ are consistent with the CB geometry.  Why is there a restriction on the choice of $g\in\Inn(F)$ that can be gauged as part of our discrete symmetry?  Relatedly, why does discrete gauging by subgroups $\G\subset\Inn(F)$, which commute with $\cN=2$ supersymmetry and leave the CB invaraint, seem not to be consistent with $\cN=2$ supersymmetry, as argued in section 4.2 of \cite{paper1}?

\end{itemize}

\acknowledgments

It is a pleasure to thank O. Aharony, M. Esole, P. Esposito, I. Garc\'ia-Etxebarria, D. Kulkarni, M. Lemos, P. Liendo, M. Lotito, Y. L\"u, L. Rastelli, D. Regalado, Y. Tachikawa, and R. Wijewardhana for helpful comments and discussions. This work was supported in part by DOE grant DE-SC0011784.  MM was also partially supported by NSF grant PHY-1151392.

\begin{appendix}
\section{Quotients of CB geometries}
\label{appA}

We demonstrate how to perform the quotient of the CB geometry by the action of a discrete subgroup the $U(1)_R$ symmetry using the SW curve and one-form.  This quotient is closely related to the discussion  in the math literature of the effect of a base change on the fiber of an elliptic surface at a ramification point of the base change; see, e.g., table 3 of \cite{SchuttShioda09}.  We illustrate with two sets of examples; all other cases follow similarly.

\subsection{$\Z_2$ quotient of the $I_{2n}$ geometry.}\label{A1}

The SW curve and one form for a scale-invariant $I_{2n}$ geometry are given by
\begin{align}\label{I2nSW}
y^2 = (x+1)(x^2 + \L^{-2n} u^{2n}), 
\qquad
\l = u \frac{dx}{y}.
\end{align}
Since the periods of the one form compute masses, it follows that $x$ and $y$ have mass dimension 0, and $u$ and $\L$ have mass dimension 1.  $\L$ is the strong coupling (or Landau pole) scale of the corresponding IR-free theory, and $u$ is the complex coordinate on the CB.

Since the power of $u$ is even, the curve is invariant under a $\Z_2$ generated by $u\mapsto -u$ leaving $x$, $y$, and $\L$ invariant.  If we orbifold the CB by this $\Z_2$ action, the complex coordinate of the resulting daughter CB is $\til u = u^2$.  The resulting curve is of Weierstrass ($y^2=x^3+\cdots$) form, but has a non-canonical SW one-form, $\l = \sqrt{\til u}\, dx/y$.  Changing variables as $y = \a^{-3} \til y$, $x = \a^{-2} \til x$ for arbitrary $\a$ preserves the Weierstrass form of the curve, and by choosing $\a$ appropriately, we can bring $\l$ to canonical form.  The unique $\a$ which does this is $\a = \sqrt{\til u}$, giving the daughter curve and one form
\begin{align}\label{In*SW}
\til y^2 =  (\til x + \til u) ( \til x^2 + \L^{-2n} \til u^{2n+2}),
\qquad
\l = \til u \frac{d\til x}{\til y},
\end{align}
which describe an $I_n^*$ singularity.  Note that now the mass dimensions of the new coordinates are $\Delta(\til u)=\Delta(\til x)=2$ and $\Delta(\til y)=3$.

The maximal mass deformation of \eqref{I2nSW} is \cite{paper2}
\begin{align}\label{I2nSWm}
y^2 = (x+1)\left(x^2 + \L^{-2n} \left[u^{2n}+ M_1 u^{2n-1} + M_2 u^{2n-2} + \cdots + M_{2n} \right] \right),
\end{align}
where the subscripts of the $M_a$ deformation parameters record their mass dimensions: $\D(M_a)=a$.  The mass deformation parameters are homogeneous polynomials in the linear mass parameters invariant under the Weyl group of the flavor symmetry.  This Weyl group is uniquely determined by the spectrum of dimensions of the mass parameters.  In this case, the spectrum is $\{1, 2, 3, \ldots, 2n\}$ which identifies the flavor Weyl group as Weyl$(U_1A_{2n-1})=$ Weyl$(U(2n))$.  Thus the (maximal) flavor symmetry of the theory corresponding to the deformation \eqref{I2nSWm} of the $I_{2n}$ singularity is $U(2n)$.  See \cite{paper1, paper2} and especially \cite{allm1602} for more details on how the flavor symmetry is inferred from the SW curve.

Now, in order for the deformed curve \eqref{I2nSWm} to be invariant under the $\Z_2$ action $u\mapsto -u$, all the odd-dimension mass deformation parameters must be set to zero, since they multiply odd powers of $u$.  Thus the resulting daughter CB geometry only has deformation parameters with a spectrum of dimensions $\{2, 4, \ldots, 2n\}$ corresponding to Weyl group Weyl$(B_n) =$ Weyl$(C_n)$, implying that the flavor symmetry algebra of the daughter theory is $\ff' = BC_n$, i.e., either $B_n$ or $C_n$.

Note that we have only discussed the maximal mass deformation of the $I_{2n}$ singularity, i.e., the one with flavor symmetry $U(2n)$.  This is the generic mass deformation of the corresponding $U(1)$ gauge theory with $2n$ charge $\pm1$ hypermultiplets.  There are many other $U(1)$ gauge theories with hypermultiplets with different charges giving the same $I_{2n}$ singularity in the zero-mass limit.  Examples of such theories appear in table \ref{Geometries}.  They correspond to geometries given by ``sub-maximal" deformations of the $I_{2n}$ singularity, with fewer mass parameters and with a different spectrum of dimensions.  A similiar $\Z_2$ orbifolding of the CB geometry can be done for these submaximal deformations, giving the results described in section \ref{U1matt}.  See \cite{paper1, paper2} for a fuller discussion of sub-maximal mass deformations.  

\subsection{$\Z_2$ and $\Z_3$ quotients of $\cN=4$ $I_0^*$ geometries.}\label{A2}

There are two different forms for the SW curve for the $\cN=4$ $\SU(2)$ SYM theory with $\cN=2$-preserving mass deformations.  As explained in \cite{paper1, paper2}, they correspond to the $I_0^* \to \{ {I_2}^3\}$ and the $I_0^* \to \{{I_1}^2, I_4\}$ deformation patterns.  In this paper we refer to them as the $I_2$-series and $I_4$-series curves, respectively.  We will discuss them in turn.

\subsubsection{Quotients of the $I_2$-series $I_0^*$ geometry}

The SW curve of the $I_2$-series $I_0^*$ geometry is given by \cite{sw2}
\begin{align}\label{I0*sw}
y^2 = \prod_{j=1}^3 (x - e_j u - e_j^2 M_2)
\end{align}
with canonical one-form $\l=u dx/y$ for $M_2=0$.  Here $e_j(\t)$ are modular forms of the marginal coupling which satisfy $\sum_j e_j=0$.  

\paragraph{$\Z_2$ quotient.}

The discriminant of the right side of \eqref{I0*sw} with respect to $x$ is proportional to $\prod_j (u-e_j M_2)$.  So only for the values of the coupling where one of the $e_j=0$ is there a $\Z_2$ symmetry on the CB.  Choose, say, $e_1=-e_3=1$ and $e_2=0$ to find the $\Z_2$-symmetric curve
\begin{align}\label{}
y^2 = x^3 - 2 x^2 M_2 - x(u^2-M_2^2)
\end{align}
with discriminant Disc$_x = 4 u^2 (u^2-M_2^2)^2$, indicative of the expected symmetrically placed $I_2$ singularities at $u=0$ and $u=\pm M_2$; see figure \ref{fig3}(c).

Now mod out by the $\Z_2$ on the CB by replacing $u$ with $\til u := u^2$, and rescaling $x$ and $y$ so that the Weierstrass form of the curve and canonical form of the 1-form are preserved.  The unique rescaling which does this is $\til x := \til u x$ and $\til y := \til u^{3/2} y$, giving a new curve
\begin{align}\label{III*toI1*I2}
\til y^2 = \til x^3 - 2 \til u M_2 \til x^2 - \til u^2 (\til u-M_2^2) \til x.
\end{align}
When $M_2=0$ limit this describes a $III^*$ Kodaira singularity.  For $M_2\neq0$, its discriminant is Disc$_x=4\til u^7(\til u-M_2^2)^2$.  As $\til u\to0$, the right side of \eqref{III*toI1*I2} becomes $\til x^3 - 2 \til u M_2 \til x^2 + \til u^2 M_2^2 \til x$ which is a singularity of $I_n^*$ type.  Since the discriminant has a factor of $\til u^7$, it must in fact be of $I_1^*$ type.  At the other singular fiber, $\til u = M_2^2$, the right side of \eqref{III*toI1*I2} becomes $\til x^2(\til x-2M_2^3)$ which has a double zero, so is of $I_n$ type.  Since the discriminant has a factor of $(\til u-M_2^2)^2$, it must in fact be of $I_2$ type.  Thus we have shown that the $\Z_2$ orbifold of the $I_0^*\to\{{I_2}^3\}$ geometry gives a curve \eqref{III*toI1*I2} which describes a $III^*\to\{I_1^*,I_2\}$ deformation pattern.

\paragraph{$\Z_3$ quotient.}

Since the discriminant of \eqref{I0*sw} is $\propto \prod_j (u-e_j M_2)$, only for the values of the coupling where the $e_j$ are the three cube roots of unity is there a $\Z_3$ symmetry on the CB.  For that coupling the curve becomes
\begin{align}\label{}
y^2 = x^3 - 3 u M_2 x - (u^3+M_2^3)
\end{align}
with discriminant Disc$_x = -27 (u^3-M_2^3)^2$, indicative of the expected $\Z_3$-symmetrically placed $I_2$ singularities; see figure \ref{fig3}(e).

Now mod out by the $\Z_3$ on the CB by replacing $u$ with $\til u := u^3$, and rescaling $x$ and $y$ so that the Weierstrass form of the curve and canonical form of the 1-form are preserved.  The unique rescaling which does this is $\til x := \til u^{4/3} x$ and $\til y := \til u^2 y$, giving a new curve
\begin{align}\label{II*toIV*I2}
\til y^2 = \til x^3 - 3 \til u^3 M_2 \til x - \til u^4 (\til u+M_2^3) .
\end{align}
When $M_2=0$ this describes a $II^*$ Kodaira singularity.  For $M_2\neq0$, its discriminant is Disc$_x=-27\til u^8(\til u-M_2^3)^2$.  As $\til u\to0$, the right side of \eqref{II*toIV*I2} becomes $\til x^3 - 3 \til u^3 M_2 \til x - \til u^4 M_2^3$ which is a singularity of $IV^*$ type.  At the other singular fiber, $\til u = M_2^3$, the right side of \eqref{II*toIV*I2} becomes $(\til x+M_2^5)^2(\til x-2M_2^5)$ which has a double zero, so is of $I_n$ type.  Since the discriminant has a factor of $(\til u-M_2^3)^2$, it must in fact be of $I_2$ type.  Thus we have shown that the $\Z_3$ orbifold of the $I_0^*\to\{{I_2}^3\}$ geometry gives a curve \eqref{II*toIV*I2} which describes a $II^*\to\{IV^*,I_2\}$ deformation pattern.

\subsubsection{Quotient of the $I_4$-series $I_0^*$ geometry}

The SW curve of the $I_4$-series $I_0^*$ geometry was found in \cite{paper2} to be given by
\begin{align}\label{I0*C1sigM}
Y^2 &= X^3
-\frac13 X [ U^2 (1+3\a^2) + 8 U M_2 \a^2 + 4 M_2^2 \a^4]
\\
&\qquad\text{}
-\frac2{27} \left[U^3 (9\a^2-1)
+3 U^2 M_2 \a^2 (5+3\a^2)
+24 U M_2^2 \a^4
+8 M_2^3 \a^6 \right],
\nonumber
\end{align}
with one-form $\l=U\, dX/Y$ at $M_2=0$.
Here $U$ is the CB parameter and $\a$ is the marginal coupling.  The curve's discriminant is $4 \a^2 (\a^2-1)^2 U^4 (U^2 + 2U M_2 + \a^2 M_2^2)$, which indicates weak coupling singularities at $\a=0,\pm1$, an $I_4$ singular fiber at $U=0$ and a pair of $I_1$ fibers at the roots of $U^2 + 2U M_2 + \a^2 M_2^2$.  

The $\Z_2$-symmetric configuration, shown in figure \ref{fig3}(a), is therefore only realized at $\a=\infty$.  This limit of the curve is accessed by defining rescaled coordinates $u:=\a^{-1} U$, $x:=\a^{-2} X$, and $y:=\a^{-3} Y$.  (This rescaling leaves the Weierstrass form of the curve and the canonical one-form unchanged.)  In terms of these new coordinates, the $\a\to\infty$ limit of the curve becomes
\begin{align}\label{}
y^2 = x^3 -\frac13 (3 u^2 + 4 M_2^2)x - \frac2{27} (9u^2 + 8 M_2^2)M_2,
\end{align}
with discriminant $4u^4(u^2+M_2^2)$, showing the expected $\Z_2$ symmetry.

Now mod out by the $\Z_2$ on the CB by replacing $u$ with $\til u := u^2$, and rescaling $x$ and $y$ so that the Weierstrass form of the curve and canonical form of the 1-form are preserved.  The unique rescaling which does this is $\til x := \til u x$ and $\til y := \til u^{3/2} y$, giving a new curve
\begin{align}\label{III*toI2*I1}
\til y^2 = \til x^3 -\frac13 \til u^2 (3 \til u + 4 M_2^2) \til x -
\frac2{27} \til u^3 (9\til u + 8 M_2^2)M_2 .
\end{align}
When $M_2=0$ this describes a $III^*$ Kodaira singularity.  For $M_2\neq0$, its discriminant is Disc$_x=4\til u^8(\til u + M_2^2)$.  As $\til u\to0$, the right side of \eqref{III*toI2*I1} becomes $\til x^3 - (4/3) \til u^2 M_2^2 \til x - (16/27) \til u^3 M_2^3$ which is a singularity of $I_n^*$ type.  Since the discriminant has a factor of $\til u^8$, it must in fact be of $I_2^*$ type.  At the other singular fiber, $\til u = -M_2^2$, the right side of \eqref{III*toI2*I1} becomes $\propto (3\til x+M_2^3)^2(3\til x-2M_2^3)$ which has a double zero, so is of $I_n$ type.  Since the discriminant has a zero of multiplicity one at $\til u =-M_2^2$, it must in fact be of $I_1$ type.  Thus we have shown that the $\Z_2$ orbifold of the $I_0^*\to\{{I_2}^2,I_4\}$ geometry gives a curve \eqref{III*toI2*I1} which describes a $III^*\to\{I_2^*,I_1\}$ deformation pattern.

%
%

\end{appendix}

\bibliographystyle{JHEP}

\end{document}